%% file: wiener_filter_short_optical_links.tex
\documentclass[journal,comsoc]{IEEEtran}

\usepackage{cite}
\usepackage{amsmath,amssymb,amsfonts}
\usepackage{algorithmic}
\usepackage{graphicx}
\usepackage{textcomp}
\usepackage{xcolor}

\usepackage[cmintegrals]{newtxmath}

\usepackage{tikz}

\usepackage{hyperref}

\ifCLASSINFOpdf
\else
\fi

\usepackage{amsmath}
\input{includes/includes.tex}

\input{includes/definitions.tex}

\usepackage{graphicx}
\usepackage{caption}
\usepackage{subcaption}
\usepackage{url}

\usepackage{etoolbox}

\begin{document}
\title{\LARGE Wiener Filter for Short-Reach Fiber-Optic Links}
\author{Daniel~Plabst,~%
        Francisco Javier Garc\'{i}a G\'{o}mez,\\~%
        Thomas Wiegart,~\IEEEmembership{Graduate Student Member,~IEEE,}\\
        Norbert~Hanik,~\IEEEmembership{Senior Member,~IEEE}%
\thanks{Date  of  current  version  \today. } %
\thanks{Daniel~Plabst, Francisco Javier Garc\'{i}a G\'{o}mez, Thomas Wiegart and Norbert~Hanik are with the Institute for Communications Engineering (LNT/L\"UT), Technical University of Munich, Munich, Germany (e-mail: d.plabst@tum.de, javier.garcia@tum.de, thomas.wiegart@tum.de, norbert.hanik@tum.de). Francisco Javier Garc\'{i}a G\'{o}mez's  work  was  supported by the German Research Foundation under Grant KR 3517/8-2.}}

\markboth{}{}%

\newcommand\copyrighttext{%
  \footnotesize \textcopyright 2020 IEEE. Personal use of this material is permitted.
  Permission from IEEE must be obtained for all other uses, in any current or future
  media, including reprinting/republishing this material for advertising or promotional
  purposes, creating new collective works, for resale or redistribution to servers or
  lists, or reuse of any copyrighted component of this work in other works,
  DOI: 10.1109/LCOMM.2020.3006921
  }
\newcommand\copyrightnotice{%
\begin{tikzpicture}[remember picture,overlay]
\node[anchor=south,yshift=5pt] at (current page.south) {\parbox{\dimexpr\textwidth-\fboxsep-\fboxrule\relax}{\copyrighttext}};
\end{tikzpicture}%
}

\makeatletter
\patchcmd{\@maketitle}
  {\addvspace{0.5\baselineskip}\egroup}
  {\addvspace{-1.0\baselineskip}\egroup}
  {}
  {}
\makeatother

\maketitle

\begin{abstract}
\input{includes/0_abstract.tex} %
\end{abstract}%


\IEEEpeerreviewmaketitle
\copyrightnotice

\input{includes/1_introduction.tex}

\input{includes/2_cont_system_model.tex}
\input{includes/3_disc_system_model.tex}
\input{includes/4_lmmse.tex}

\input{includes/5_numerical_results.tex}
\input{includes/6_conclusion.tex}

\section*{Acknowledgment}
\noindent The authors wish to thank Prof. Gerhard Kramer and Prof. Amine Mezghani for useful suggestions and discussions.

\ifCLASSOPTIONcaptionsoff
  \newpage
\fi

\bibliographystyle{IEEEtran}
\vspace*{-0pt}
\bibliography{IEEEabrv,imdd}
\end{document}

%% file: includes/includes.tex
\usepackage{trfsigns}
\usepackage{upgreek}

\usepackage{tikz}
\usetikzlibrary{calc,shapes,arrows,fit,decorations.markings,patterns,fadings,positioning}

\tikzstyle{sum} = [thick,draw, circle,inner sep=-1pt,outer sep=0pt,font=\normalsize]

\pgfmathsetmacro{\blockwidth}{30}
\pgfmathsetmacro{\blockheight}{17}

\tikzstyle{comblock} = [thick,black,fill=white,draw=black, rectangle,minimum height=\blockheight pt,minimum width=\blockwidth pt]
\tikzstyle{comtriag} = [regular polygon, regular polygon sides=3,
              draw=black, black,  text width=0.90em,
              inner sep=0.9mm, outer sep=0mm,
              shape border rotate=-90]

\tikzstyle{input} =  [coordinate]
\tikzstyle{output} = [coordinate]

\tikzset{%
Double/.style={%
    to path={%
      ($(\tikztostart)!2pt!90:($(\tikztotarget)!5pt!(\tikztostart)$)$) -- ($($(\tikztotarget)!2.3pt!(\tikztostart)$)!2pt!270:(\tikztostart)$)
      ($(\tikztostart)!2pt!270:($(\tikztotarget)!4pt!(\tikztostart)$)$) -- ($($(\tikztotarget)!2.3pt!(\tikztostart)$)!2pt!90:(\tikztostart)$)
      ($($(\tikztotarget)!5pt!(\tikztostart)$)!5pt!90:(\tikztostart)$)
       .. controls
          ($($(\tikztotarget)!3pt!(\tikztostart)$)!3pt!90:(\tikztostart)$) and
          ($(\tikztotarget)!0.1pt!(\tikztostart)$)
       .. (\tikztotarget)
       .. controls
          ($(\tikztotarget)!0.1pt!(\tikztostart)$) and
          ($($(\tikztotarget)!3pt!(\tikztostart)$)!3pt!270:(\tikztostart)$)
       ..
     ($($(\tikztotarget)!5pt!(\tikztostart)$)!5pt!270:(\tikztostart)$)
    }
  }
}

\usepackage[exponent-product = \times]{siunitx}
\usepackage{graphicx}

\usepackage{cleveref}
\usepackage{mathtools}
\usetikzlibrary{patterns,fadings}

\usepackage{pgfplots}
\pgfplotsset{compat=1.14}

%% file: includes/definitions.tex
\DeclareMathOperator*{\h}{H}
\DeclareMathOperator*{\T}{T}
\DeclareMathOperator*{\diag}{diag\mkern-2.8mu}

\DeclareMathOperator{\E}{\mathbb{E}}
\DeclareMathOperator{\sinc}{sinc}
\DeclareMathOperator{\trace}{tr}

\newcommand{\Real}[1]{\ensuremath{\Re\left\{#1\right\}}}
\newcommand{\Imag}[1]{\ensuremath{\Im\left\{#1\right\}}}

\newcommand{\GaussianPDF}[2]{\ensuremath{\mathcal{N}\big( {#1}, {#2} \big)}}

\newcommand{\moment}[1]{\ensuremath{\mathrm{\mu}_{4}}}

\newcommand{\uarg}{\ensuremath{{\bm{\cdot}}}}

\newcommand{\ti}{\ensuremath{{\nu}}}

\newcommand{\tip}{\ensuremath{{\kappa}}}

\newcommand{\dimZ}[1]{\ensuremath{\in \mathbb{Z}^{#1}}}
\newcommand{\dimC}[1]{\ensuremath{\in \mathbb{C}^{#1}}}
\newcommand{\dimR}[1]{\ensuremath{\in \mathbb{R}^{#1}}}
\newcommand{\dimRP}[1]{\ensuremath{\in \mathbb{R}^{#1}_+}} %
\newcommand{\covm}[2]{\ensuremath{\boldsymbol{\mathbf{#1}}_{{#2}}}}
\newcommand{\mean}[1]{\ensuremath{\boldsymbol{\mathbf{\upmu}}_{{#1}}}}
\newcommand{\mident}[1]{\ensuremath{\mathrm{\textbf{I}}_{#1}}}
\newcommand{\mnull}[1]{\ensuremath{\mathrm{\textbf{0}}_{#1}}}
\newcommand{\mones}[1]{\ensuremath{\mathrm{\textbf{1}}_{#1}}}
\newcommand{\unitvec}[1]{\ensuremath{\mathrm{\textbf{e}}_{#1}}}

\newcommand{\lmmse}{\ensuremath{\bm{g}}}
\newcommand{\Tsym}{\ensuremath{T_\text{s}}}
\newcommand{\meanlmmse}{\ensuremath{g_m}}
\newcommand{\combchannel}{\ensuremath{{\psi}}}
\newcommand{\combchannelmat}{\ensuremath{{\mathbf{\Psi}}}}

\newcommand{\WF}{\ensuremath{\mathrm{WF} }}
\newcommand{\WFlin}{\ensuremath{\widetilde{\mathrm{WF}}}} %

\newcommand{\bm}[1]{\ensuremath{\boldsymbol{\mathbf{#1}}}}

\definecolor{mycolor2}{RGB}{196,7,27}
\definecolor{mycolor1}{RGB}{0,101,189}
\definecolor{bblack}{RGB}{088,088,090}
\definecolor{oorange}{RGB}{255,180,000}
\definecolor{mycolor3}{RGB}{255,180,000}
\definecolor{mycolor4}{HTML}{3A7E55}
\definecolor{mycolor5}{HTML}{3A7E55}

\newcommand\numberthis{\addtocounter{equation}{1}\tag{\theequation}}

%% file: includes/0_abstract.tex
Analytic expressions are derived for the Wiener filter (WF), also known as the linear minimum mean square error (LMMSE) estimator, for an intensity-modulation/direct-detection (IM/DD) short-haul fiber-optic communication system. 
The link is purely dispersive and the nonlinear square-law detector (SLD) operates at the thermal noise limit. 
The achievable rates of geometrically shaped PAM constellations are substantially increased by 
taking the SLD into account as compared to a WF that ignores the SLD. 

%% file: includes/1_introduction.tex
\section{Introduction}
\IEEEPARstart{S}{hort}-reach fiber-optic communications systems, e.g., for data-center interconnects,
usually use transceivers based on intensity-modulation (IM) and direct detection (DD) (e.g.,~\cite{8649641,7779079}).
Compared to coherent transceivers, IM/DD transceivers offer lower power consumption and hardware complexity, smaller form factors and hence reduced overall costs~\cite{8649641,8259239}. 
To further reduce cost and complexity, short-link communication systems are usually operated without optical amplification and dispersion compensating fiber and require signal distortions to be compensated digitally at the receiver. 

In short-reach communication systems, inter-symbol-interference (ISI) caused by chromatic dispersion (CD) is the limiting effect~\cite{7779079,8259239,7341665}. CD is described by a complex-valued impulse response. %
A DD receiver consists of a photodiode which measures the intensity of the imminent electrical field, and hence discards phase information. This complicates CD removal. Due to the absence of amplifiers on short-reach links, the square-law detector (SLD) is the only noise source. Since the receive signal is significantly attenuated, the SLD is assumed to operate at the thermal noise limit and adds white Gaussian noise to the intensity measurements~\cite[P.~154]{AgrawalFourthEdFiberOptics}.  

Common CD equalizers include linear feed-forward equalization (FFE) or non-linear methods like decision feedback equalizing (DFE), Volterra series based equalization, and neural network based equalization (e.g.,~\cite[Sec. IV]{8259239},~\cite{Karanov:18}).
In this paper, we consider a linear equalizer, namely the minimum mean square error (MMSE) estimator, also known as the Wiener filter (WF). We derive analytic expressions for the WF coefficients for short-reach IM/DD systems. Due to small transmit signal powers, the Kerr nonlinearity of the link can be neglected~\cite[P.~65]{AgrawalFourthEdFiberOptics} and the link is purely dispersive. 

In~\cite{abs-1802-00432} the authors compute the WF assuming either real-valued Gaussian transmit symbols and a real-valued channel matrix or circularly symmetric complex Gaussian transmit symbols and a complex-valued channel matrix. We consider real-valued transmit symbols, originating from  any symmetric probability density function (PDF), and a complex-valued channel matrix and  extend the expressions from~\cite{abs-1802-00432}.
\newline\indent
\textit{Notation}: Bold letters indicate vectors and matrices, non-bold letters express scalars. For a matrix $\bm{A}$, we denote complex conjugate, transpose and Hermitian transpose by
$\bm{A}^*$, $\bm{A}^{\T}$ and $\bm{A}^{\h}$, respectively.
The Hadamard product and the trace operator are expressed by $\left(\circ\right)$ and $\trace{\left(\uarg\right)}$, respectively.
By $\diag{\left(\bm{a}\right)}$ we denote a square matrix with the vector $\bm{a}$ on its main diagonal, while $\diag{\left(\bm{A}\right)}$ outputs the main diagonal of $\bm{A}$ as a column vector.
The $N \times N$ identity matrix is written as $\mident{N \times N}$, while the $N \times 1$ all-ones and all-zeros vector are referred to by $\mones{N}$, $\mnull{N}$, respectively.
The mean of a random vector $\bm{a}$ is expressed by $\mean{\bm{a}}$ and the covariance matrix of two random vectors $\bm{a}$, $\bm{b}$ is denoted by $\covm{C}{\bm{a}\bm{b}}$.
By $\unitvec{i} $, we denote the canonical unit (column) vector of appropriate dimensions, with all entries equal to zero, except the $i$-th ($0$-based indexing). %
Dirac's delta is expressed by $\delta(t)$ and we use $\langle a(t), a(t)\rangle = \int_{-\infty}^{\infty} \big|\, a(t) \big|^2 \mathrm{d}t $ to indicate the energy of $a(t)$.
The sinc function is defined as $\sinc(\pi s) = \sin(\pi s)/(\pi s)$.
The Nyquist ISI-free property of a pulse $g(t)$ with symbol period $\Tsym$ reads $\left. g(t)\right|_{t=k\Tsym} = g(0)\delta[k]$, $\forall k \dimZ{}$. 
A pulse $g(t)$ has the $\sqrt{\text{Nyquist}}$ property, if $g(t)\,*\,g^*(-t)$ has the Nyquist property.
The Fourier pair $a(t) = \mathcal{F}^{-1}\big\{ A(f) \big\}$ and $A(f) = \mathcal{F}\left\{ a(t) \right\}$, is denoted by $a(t)$ \laplace\, $A(f)$. By $f(\mathcal{A})$ we denote that the function $f(\uarg)$ is applied element-wise to the set $\mathcal{A}$, i.e., $f(\mathcal{A}) = \left\lbrace f(a)\,|\,  a \in \mathcal{A}  \right\rbrace$. By $\Real{\bm{A}}$ and $\Imag{\bm{A}}$ we denote element-wise real and imaginary part of the complex-valued matrix $\bm{A}$, respectively.

%% file: includes/2_cont_system_model.tex
\section{System Model}
\label{sec:system_model}
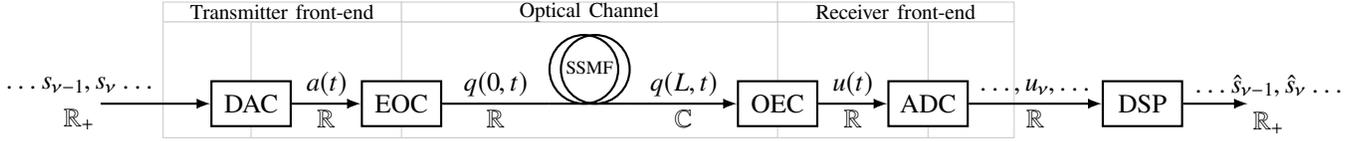
\begin{figure*}[!ht]
  \centering
  \input{Figures/system_model_flow_graph.tex}
  \vspace*{-9pt}
  \caption{Signal flow graph of a single-carrier IM/DD communication system.}
  \label{fig:Signal_Flow_Graph_IM_DD}
  \vspace*{-10pt}
\end{figure*}
\subsection{Transmitter Front-End}
\label{sec:transmitter_front_end}
In Fig.~\ref{fig:Signal_Flow_Graph_IM_DD}, the transmitter is fed with \textit{positive} and \textit{real-valued}, random, discrete-time data \textit{symbols} $s_\ti$, where $\ti$ is the discrete time index.
We have $s_\ti \!\in\! \mathcal{S}$ and  modulation alphabet $\mathcal{S}$. %
\subsubsection{Digital-to-Analog Converter (DAC)}
For \textit{ideal} digital to analog conversion of the $s_\ti$, the DAC performs pulse shaping with symbol time $\Tsym$. The continuous-time DAC output $a(t)$ reads
\begin{align}
    a(t) = s(t) * g_\text{tx}(t) = \sum\nolimits_{\ti = -\infty}^{\infty} s_\ti g_\text{tx}\left( t - \nu  \Tsym \right)
    \label{eq:pulseshaping_gtx}
\end{align}
with real-valued pulse-shaping filter $g_\text{tx}(t)$ and 
\begin{align}
  s(t) = \sum\nolimits_{\ti = -\infty}^{\infty}  s_\ti  \delta(t-\nu \Tsym)
  .
  \label{eq:s_t_cont}
\end{align}
\subsubsection{Electrical-Optical Converter (EOC)}
\label{sec:eoc}
We use an \textit{ideal} Mach-Zehnder Modulator (MZM) in push-pull mode~\cite[P. 19]{seimetz2009high}
as the EOC, which is shown in Fig.~\ref{fig:MZM}. The MZM modulates the amplitude of an %
electromagnetic carrier wave from a laser.
The electric field $\bar{q}(t)$ of the laser light reads
\begin{align}
  \bar{q}(t) = E_0 \cdot \e^{ -\im  \omega_c t } \;
\end{align}
with  $E_0 \!=\! \hat{E}_0\, \e^{\im \phi_{E_0}(t)} \!\in \mathbb{C}$, $\hat{E}_0 \!\triangleq\!  \big| E_0 \big|$ and angular frequency $\omega_c$.
Laser fluctuations %
\cite[P. 100]{AgrawalFourthEdFiberOptics} are neglected and we set
$\phi_{E_0}(t) \!=\! 0$. Modulating $\bar{q}(t)$ %
with $x(t)$ leads to a complex-valued bandpass signal 
at the fiber input $z \!=\! 0$~\cite[P. 19]{seimetz2009high}:
\begin{align}
  \tilde{q}(0,t)= \underbrace{\cos{ \bigg\{ x(t) \cdot \frac{\pi}{2 V_\pi} \bigg\} }}_{\text{Information signal $q(0,t)$} } \;\;\cdot \;\;\; \underbrace{\bar{q}(t) \vphantom{\left\lbrace x(t)  \frac{\pi}{2 V_\pi}\right\rbrace
   }}_{\mathclap{\text{Carrier signal}}}
  \numberthis
  \label{eq:mzm_baseband_carrier}
\end{align}
with hardware constant $V_\pi$ of the MZM and we set $V_\pi = \pi/2$.
We decompose $x(t)$ into a bias $\bar{x} = \pi/2$ and alternating signal $\tilde{x}(t)$, i.e., $x(t) = \bar{x} + \tilde{x}(t)$, and use the small angle approximation for $\cos(\uarg)$ around $\bar{x}$, i.e.,
$\cos\left(\bar{x} + \tilde{x}(t) \right) \,\!\approx\!\, - \tilde{x}(t)$. %
The approximation error is small for $\left\lvert \tilde{x}(t) \right\rvert \ll \pi/2 $. The bandpass signal $\tilde{q}(0,t)$, launched into the fiber, reads as
\begin{align}
  \tilde{q}(0,t) \approx - \hat{E}_0 \cdot \tilde{x}(t) \cdot \,\e^{-\im\omega_c t}
  \numberthis
  \label{eq:mzm_baseband_approx}
\end{align}
and we define $a(t) \triangleq - \hat{E}_0 \,\tilde{x}(t)$. Neglecting the carrier signal term from~\eqref{eq:mzm_baseband_approx}, we obtain the  real-valued baseband signal
\begin{align}
  q(0,t)=  a(t) \;
  \numberthis
  \label{eq:mzm_baseband_simplified}
  .
\end{align}
Modulating the \textit{amplitude} of the baseband electric field mo\-dulates the optical \textit{intensity} $I(z,t)$ at $z=0$: 
\begin{align}
  I(0,t)= \gamma_\text{prop} \cdot \left\lvert q(0,t) \right\rvert^2 =  \gamma_\text{prop} \cdot  a(t)^2
  \numberthis
  \label{eq:intensity_baseband}
\end{align}
with  constant $\gamma_\text{prop} \!\triangleq\! 1$. For invertible relationships between amplitude and intensity, i.e., real-valued non-negative $a(t)$, this scheme is referred to as IM~\cite[P. 20]{seimetz2009high}. Using a MZM, we require $a(t)$ to only be real-valued, but justify a non-negativity condition of the transmit symbols in Sec.~\ref{sec:tx_symbol_dist}. 
\begin{figure*}
  \begin{minipage}{.5\textwidth}
    \centering
    \input{Figures/mzm.tex}
    \vspace*{-4pt}
    \caption{A Mach-Zehnder Modulator as  EOC.}
    \label{fig:MZM}
  \end{minipage}%
  \begin{minipage}{.5\textwidth}
    \centering
    \input{Figures/oec.tex}
    \vspace*{-4pt}
    \caption{Square-law detector as OEC with noise and receive filter. }
    \label{fig:OEC}
  \end{minipage}
  \vspace{-12pt}
\end{figure*}

\subsection{Optical Channel}
The propagation of the slowly varying signal $q \!\triangleq \!q(z,t)$
is
described by the nonlinear Schr\"odinger equation~\cite[P. 65]{AgrawalFourthEdFiberOptics}
\begin{align}
  \frac{\partial q}{\partial z} &=
  - \im \frac{\beta_2}{2} \frac{\partial^2 q}{\partial t^2}
  \;\vphantom{\frac{\,}{\,}} + \im\gamma \lvert q\rvert^2 q
  \;-\frac{\alpha}{2}q
  \;\vphantom{\frac{\beta_2}{2}} + n
  \label{eq:nlse_full}
\end{align}
where $\beta_2$ is the CD coefficient,
$\gamma$ is the Kerr nonlinearity parameter,
$\alpha$ accounts for fiber-loss and $z$ is the propagated distance. 
The term $n \triangleq n(z,t)$ describes noise realizations. %
The Kerr nonlinearity can be neglected for small optical transmit powers $P_\text{tx,opt}$~\cite[P. 65]{AgrawalFourthEdFiberOptics}. %
With no amplification along the fiber, the dominant noise is added by the SLD at the receiver~\cite{8649641}. We thus set $n \!=\! 0$ and model electrical noise of the SLD in the following section.
Finally, we consider attenuation in the signal-to-noise ratio (SNR) definition at the receiver and therefore simplify~\eqref{eq:nlse_full} to a \textit{linear} differential equation
\begin{align}
  \frac{\partial q}{\partial z} &=
  - \im \frac{\beta_2}{2} \frac{\partial^2 q}{\partial t^2}
  \label{eq:nlse_full_simplified}
\end{align}
which can be solved analytically in the Fourier domain as
\begin{align}
  Q(L,\omega) &= Q(0,\omega) \cdot \e^{ +\im \frac{\beta_2}{2}\omega^2 L  }
\label{eq:nlse_simplified_fourier}
\end{align}
with $Q(\uarg,\omega) \,\Laplace\, q(\uarg,t)$, frequency response of CD  $H(L,\omega) \!\triangleq\! \e^{\! +\im \frac{\beta_2}{2}\omega^2 L  }$, 
 $H(L,\omega)  \,\Laplace\, h(L,t)$ and fiber length $L$.
\subsection{Receiver Front-End}
The receiver performs optical to electrical conversion (OEC), digitizes the signal by an ideal analog-to-digital converter (ADC) and recovers the transmitted data by DSP.
The receiver in Fig.~\ref{fig:OEC} consists of a \textit{p-i-n} SLD~\cite[P. 153]{AgrawalFourthEdFiberOptics}, with output current $r^\prime(t)$ proportional to the intensity of the impinging electrical field, i.e.,
$r^\prime(t) = \gamma_\text{pd} \cdot \left\lvert q(L,t) \right\rvert^2,$
and proportionality constant $\gamma_\text{pd} \!\triangleq \!1$~\cite[Eq. (4.1.2-3)]{AgrawalFourthEdFiberOptics}.
With no amplifiers on the fiber, $q(L,t)$ at the fiber end will be significantly attenuated compared to $q(0,t)$, which allows consideration of the receiver at the thermal noise limit~\cite[P. 154]{AgrawalFourthEdFiberOptics}. %
Therefore, $\eta^\prime(t)$ is described by a
\textit{white} Gaussian random process with  two-sided power spectral density (PSD) $\Phi_{\eta^\prime \eta^\prime}(f) = N_0/2$, autocorrelation function (ACF) $\phi_{\eta^\prime \eta^\prime}(\tau) = (N_0/2) \delta(\tau)$, time-lag $\tau$ and $N_0$ as stated in~\cite[Eq. 4.4.7]{AgrawalFourthEdFiberOptics}. With bandwidth limitation in the electrical filter $g_\text{rx}(t)$, prior to the ADC, the noise energy is finite and communication viable.

%% file: Figures/system_model_flow_graph.tex
\usetikzlibrary{decorations.markings}
\tikzset{node distance=2cm}

\pgfdeclarelayer{background}
\pgfdeclarelayer{foreground}
\pgfsetlayers{background,main,foreground}

\pgfmathsetmacro{\circledia}{13}

\tikzset{midnodes/.style = {midway,above,text width=1.5cm,align=center,yshift=-1.3em}}
\tikzset{boxlines/.style = {draw=black!20!white,}}
\tikzset{boxes/.style = {draw=black!20!white,fill=white}}

\begin{tikzpicture}[]
    \node [input, name=input] {Input};
    \node [comblock,right of=input] (dac) {DAC};
    \node [comblock,right of=dac] (eoc) {EOC};
    \node [comblock,right of=eoc,draw=none,node distance=2.5cm] (fiber_channel){};
    \node [comblock,right of=fiber_channel,node distance=2.5cm] (oec) {OEC};
    \node [comblock,right of=oec] (adc) {ADC};
    \node [comblock,right of=adc,node distance=2.85cm] (dsp) {DSP};
    \node [input, name=output, right of=dsp,node distance=1.4cm] {Output};

    \node [input, name=input_two,right of=input, node distance=0.9cm] {};

    \draw[thick] (fiber_channel) ++(-\circledia/6pt,\circledia pt) circle[radius=\circledia pt];   %
    \draw[thick] (fiber_channel) ++(\circledia/6pt,\circledia pt) circle[radius=\circledia pt];   %

    \draw[-latex,thick] (input) node[midway,left,text width=2.2cm,align=center,xshift=0.95cm](){$\ldots s_{\ti-1}, s_\ti \ldots$\\[0.2em] $\mathbb{R}_+$ } -- (dac);
    \draw[-latex,thick] (dac) --node[midnodes](){$a(t)$\\[0.2em] $\mathbb{R}$}  (eoc);
    \draw[thick] (eoc) -- node[midnodes](){$q(0,t)$\\[0.2em] $\mathbb{R}$}  (fiber_channel);
    \draw[thick] (fiber_channel.west) --node[midway,font=\scriptsize,above,yshift=0.5*\circledia pt +0.1em]{SSMF} (fiber_channel.east);
    \draw[-latex,thick] (fiber_channel) -- node[midnodes](){$q(L,t)$\\[0.2em] $\mathbb{C}$}  (oec);
    \draw[-latex,thick] (oec) -- node[midnodes](){$u(t)$\\[0.2em] $\mathbb{R}$} (adc);
    \draw[-latex,thick] (adc) -- node[midnodes](){$\ldots ,u_\ti, \ldots$\\[0.2em] $\mathbb{R}$} (dsp);
    \draw[-latex,thick] (dsp) -- (output)  node[right,text width=2.2cm,xshift=-0.95cm,align=center](){$\ldots \hat{s}_{\ti-1}, \hat{s}_\ti \ldots$\\[0.2em] $\mathbb{R}_+$};

    \begin{pgfonlayer}{background}
        \pgfmathsetmacro{\blockheight}{31}
        \draw[boxlines] (eoc) ++(0,\blockheight pt) rectangle node[midway,below,yshift=\blockheight/1.3 pt +1pt,font=\footnotesize]{} ($(oec) + (0,-\blockheight/2.4 pt)$);
        \draw[boxlines] (dac) ++(0,1.00*\blockheight pt) rectangle node[midway,below,xshift=-3.5cm,yshift=\blockheight/1.35 pt,font=\footnotesize]{} ($(adc) + (0,-\blockheight/2.4 pt)$);
        \draw[boxlines] (input_two) ++(-2pt ,1.0*\blockheight pt) rectangle node[midway,below,yshift=\blockheight/1.35pt, xshift=-5.3cm,font=\footnotesize]{} ($($(adc.east)!0.5!(dsp.west)$) + (-8pt,-\blockheight/2.4 pt)$);

        \pgfmathsetmacro{\adddist}{0}

        \draw[fill=white,draw=white] (input_two) ++(-2 pt ,1.24*\blockheight pt) rectangle node[midway]{} ($($(adc.east)!0.5!(dsp.west)$) + (-\adddist pt,1.0*\blockheight pt)$);
        \draw[boxes] (input_two) ++(-2 pt ,1.24*\blockheight pt) rectangle node[midway,font=\footnotesize]{Transmitter front-end} ($(eoc) + (-\adddist pt,1.0*\blockheight pt)$);
        \draw[boxes] (eoc) ++(\adddist pt,1.24*\blockheight pt) rectangle node[midway,font=\footnotesize]{Optical Channel} ($(oec) + (-\adddist pt,1.0*\blockheight pt)$);
        \draw[boxes] (oec) ++(\adddist pt,1.24*\blockheight pt) rectangle node[midway,font=\footnotesize]{Receiver front-end} ($($(adc.east)!0.5!(dsp.west)$) + (-8 pt,1.0*\blockheight pt)$);
    \end{pgfonlayer}
\end{tikzpicture}

%% file: Figures/mzm.tex
\usetikzlibrary{decorations.markings}
\tikzset{node distance=1.0cm}

\tikzset{boxlines/.style = {dashed,draw=red!50!white,thick}}

\pgfdeclarelayer{background}
\pgfdeclarelayer{foreground}
\pgfsetlayers{background,main,foreground}

\pgfmathsetmacro{\sepwidth}{53}
\pgfmathsetmacro{\sepwidthtwo}{38}

\begin{tikzpicture}[]
    \node [input, name=input] {Laser};
    \node [comblock,draw=none,right of=input,inner sep=1pt,node distance=1cm] (laserEMW) {$ \bar{q}(t)$};
    \node [sum,right of=laserEMW,inner sep=1pt,node distance=1.5cm] (multiplier) {$\times$};
    \node [comblock,above of=multiplier,node distance=0.9cm] (cos_block) {$\cos{\big\{ \left(\uarg\right)  \frac{\pi}{ 2 V_\pi} \big\}}$};
    \node [input, name=electrical,above of=cos_block,node distance=0.8cm,xshift=-1em] (electrical) {$x$};
    \node [comblock,draw=none, name=optical,right of=multiplier,node distance=1.5cm] (MZMoutput) {$ \tilde{q}(0,t)$};
    \node [input, name=optical,right of=MZMoutput,node distance=0.8cm] (optical) {};

    \node[left] () at (electrical) {$x(t) = \tilde{x}(t) + \bar{x}$};
    \node[left,font=\footnotesize,xshift=-2.4cm,yshift=0cm] () at (electrical) {DAC output:};
    \node[left,font=\footnotesize] () at (input) {Laser input: };
    \node[right,font=\footnotesize] () at (optical) {$\ldots$ to fiber};

    \draw[-latex,thick] (laserEMW) -- (multiplier);
    \draw[-latex,thick] (multiplier) -- (MZMoutput);
    \draw[-latex,thick] (cos_block) -- (multiplier);
    \draw[-latex,thick] (electrical) -|  (cos_block);
    \begin{pgfonlayer}{background}
        \draw[boxlines] ($( input |- cos_block) + (-\sepwidth pt,0)$) -- node[,xshift=-23pt,left,text width=3cm,font=\footnotesize,align=left]{Electrical\\[0.2em] Optical}  ($( optical.east |- cos_block) + (+\sepwidthtwo pt,0)$);
    \end{pgfonlayer}
\end{tikzpicture}

%% file: Figures/oec.tex
\usetikzlibrary{decorations.markings}
\tikzset{node distance=1.0cm}

\tikzset{scalepd/.style = {scale=0.6,transform shape}}
\usetikzlibrary{arrows, decorations.pathmorphing}
\tikzset{boxlines/.style = {densely dashed,draw=red!50!white,thick}}

\pgfdeclarelayer{background}
\pgfdeclarelayer{foreground}
\pgfsetlayers{background,main,foreground}

\begin{tikzpicture}[ decoration = {snake,   %
                    pre length=3pt,post length=7pt,%
                    }]
    \node [input, name=input] {optical};
    \node [comblock,right of=input,inner sep=1pt,minimum width=1.6cm,node distance=2.5cm,font=\footnotesize] (pd) {$\;\big| \, \uarg \, \big|^2$};
    \node [sum,right of=pd,node distance=2cm] (sumblock) {+};
    \node [comblock,right of=sumblock,inner sep=1pt,node distance=1.5cm,font=\footnotesize] (rxchain) {
    $g_\text{rx}(t)$
    };
    \node [input,right of=rxchain,node distance=1.5cm] (dac) {};
    \node[above,yshift=0.5cm,font=\footnotesize,text width=2cm,align=center] (rxchaindesc) at (rxchain) {Electrical\\ receive filter};
    \node[above,yshift=1cm] (noise) at (sumblock) {$\eta^\prime(t)$};
    \node[above right,xshift=-0.8cm,text width=2cm,align=left] () at (dac) {$u(t)$};
    \node[right,yshift=0cm,font=\footnotesize,text width=2cm,align=left] () at (dac) {$\ldots$ to ADC};
    \path[draw=black, decorate,-latex] (input) --node[midway,above,yshift=0.1cm]{$q(L,t)$} (pd);
    \draw[-latex] (noise) --(sumblock);
    \draw[-latex] (pd) -- node[above]{$r^\prime(t)$}(sumblock);
    \draw[-latex] (sumblock) --  node[above]{$r(t)$}(rxchain);
    \draw[-latex] (rxchain) -- (dac);
    \begin{pgfonlayer}{background}
        \draw[boxlines] ($( pd |- noise) + (0,+07pt)$) -- node[left,yshift=1.2cm,text width=1.1cm,rotate=90,font=\footnotesize,align=right,]{Optical\\[0.2em] Electrical}  ($( pd |- noise) + (0,-50pt)$);
    \end{pgfonlayer}
\end{tikzpicture}
\vspace{-9pt}

%% file: includes/3_disc_system_model.tex
\section{Discrete-Time System Model}
\label{sec:discrete_time_system_model}
\subsection{Nyquist System}
\subsubsection{Linear Case}
\begin{figure*}[htb]
  \centering
  \input{Figures/detailed_system_model.tex}
  \vspace{-14pt}
  \caption{End-to-end baseband model of the IM/DD lightwave communication system.}
  \label{fig:Detailed_Flow_Graph_IM_DD}
  \vspace{-8pt}
\end{figure*}
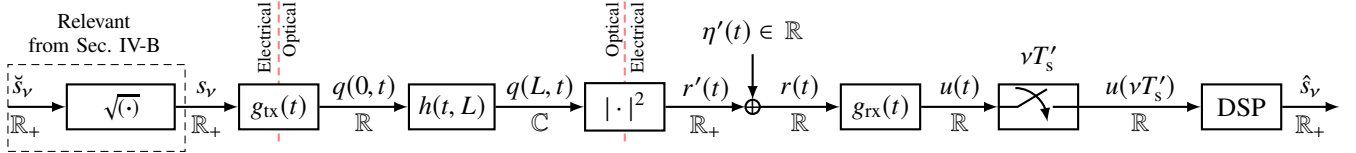

A \textit{linear} communication system (Fig.~\ref{fig:Detailed_Flow_Graph_IM_DD} \textit{without} SLD) with $\Tsym$-spaced sampling at the receiver has zero ISI, maximum SNR, and additive white Gaussian noise (AWGN) at the sampling times for
$g(t) \! =\! g_\text{tx}(t) * h(L,t) $ being $\sqrt{\text{Nyquist}}$  and $g_\text{rx}(t)$  the matched filter~\cite[PP. 175]{gallager2008principles}.

\subsubsection{Nonlinear Case}
Consider the SLD, zero CD and a single symbol $s_\ti$ being pulse-shaped and transmitted, which requires a $\sqrt{\text{Nyquist}}$ $r^\prime(t)$  and $g_\text{rx}(t)$ as the  matched filter. However,~\cite{4132995} showed that nonnegative, \textit{bandlimited} $\sqrt{\text{Nyquist}}$ pulses do not exist and that practical nonnegative $\sqrt{\text{Nyquist}}$ pulses are time-limited to $\Tsym$.
However, zero ISI and AWGN at the receiver at multiples of $\Tsym = 1/B$ are still viable, when, e.g., choosing $g_\text{tx}(t)$ as a \textit{sinc} pulse shaping filter with bandwidth $B$, and $g_\text{rx}(t)$ as a \textit{sinc} filter with bandwidth $2B$. Since \textit{sinc} filters fulfill Nyquist and $\sqrt{\text{Nyquist}}$ criterion, the zero-ISI condition at $\Tsym$-spaced sampling is met and the receiver noise samples are uncorrelated, which we show in the following and Sec.~\ref{sec:tx_symbol_dist}.
\subsubsection{Nonlinear and Dispersive Case} With CD, zero ISI at multiples of $\Tsym$ 
is not viable, as a real-valued $g_\text{tx}(t)$ cannot  form an ISI-free filter with the complex-valued CD $h(L,t)$. Thus, we formulate an optimization problem for the remaining ISI.

\subsection{Bandlimited Sampling Receiver}
We choose $g_\text{tx}(t) \!=\! \sinc{\left( B \pi  t \right)} \,\laplace\, G_\text{tx}(\omega)$ as the band\-limited transmitter pulse shaping filter, with two-sided bandwidth $B \!=\! 1/\Tsym $.
With the number of transmit symbols $V$, the average \textit{optical} transmit power $P_\text{tx,opt}$ per symbol reads
\begin{align}
  P_\text{tx,opt}  \!\triangleq\! \lim_{V \rightarrow \infty}
  \frac{1}{V \Tsym}  \E_s \left[ \langle a(t),\, a(t)\rangle \right] =  %
      \sigma_\mathrm{s}^2 + \mu_\mathrm{s}^2
\end{align}
We  therefore adjust the average fiber launch power with the symbol variance $\sigma_\mathrm{s}^2$ and mean $\mu_\mathrm{s}$.
The receive filter $g_\text{rx}(t)$ has unit frequency gain and sets the receiver bandwidth to $2 B$, i.e.,
\begin{align}
  g_\text{rx}(t) \!=\! 2B \!\cdot\!\sinc\left(2B \pi t \right)\;\laplace\; G_\text{rx}(f) \! =\!
  \begin{cases}
    1, \, \text{if }\left\lvert f \right\rvert \leq  B\\
    0, \, \text{otherwise}
  \end{cases}\hspace*{-10pt}
\end{align}
given that the bandwidth from $q(L,t)$ to $r^\prime(t)$ is doubled by the $\left| \,\uarg\,\right|^2$-operation of the SLD.
Thus, the receiver is a bandlimited sampling receiver, sampling $u(t)$ at $N_\text{os} \times B$, where $N_\text{os} \geq 2$ to avoid aliasing.
The noise energy in $u(t)$ is thus finite. With real-valued transmit symbols, the PSD $\Phi_{\eta\eta}(f)$ and ACF $\phi_{\eta\eta}(\tau)$ of the bandlimited \textit{real-valued} noise read~\cite[P. 233]{gallager2008principles},
\begin{align}
  \Phi_{\eta\eta}(f) \!=\! \begin{cases}
   \frac{N_0}{2}, \;\; \text{for}\; \left\lvert f \right\rvert \leq B \!\\
   \,0, \;\;\;\; \text{otherwise} \!
  \!\end{cases} \hspace{-8pt} \Laplace \; \phi_{\eta\eta}(\tau) = \frac{N_0}{2} \phi_{g_\text{rx}g_\text{rx}}(\tau)  %
\end{align}
with $\phi_{\eta\eta}(\tau) \!=\! \phi_{\eta^\prime\eta^\prime}(\tau) \,\!*\!\, \phi_{g_\text{rx}g_\text{rx}}(\tau)$, $\phi_{g_\text{rx}g_\text{rx}}(\tau) \!=\! 
 g_\text{rx}(\tau)$.%
\subsection{Discrete-Time Formulation}
Sampling at $t = \nu  \Tsym^\prime$ with $\Tsym^\prime = \Tsym/N_\text{os}$ and $N_\text{os} \geq 2$ gives
\begin{align}
  u(\kappa\Tsym^\prime )  \triangleq u[\tip] \!=\!  r^\prime[\tip] + \eta[\tip]
  \label{eq:uoft_nyquist_sampling}
\end{align}
where the  discrete-time instant is $\tip \!\in\! \mathbb{Z}$ and $\eta[\tip]$ is zero-mean real-valued Gaussian noise with ACF
 $\phi_{\eta\eta}[\tip]  \triangleq \phi_{\eta\eta}(\tau\! =\! \ti  \Tsym^\prime) $.
For $N_\text{os}\!=\! 2$, the noise is \textit{white}, i.e., $\phi_{\eta\eta}[\tip] = N_0 B \delta[\tip]$
and therefore $\eta[\tip] \!\sim\! \GaussianPDF{0}{\sigma_\mathrm{\eta}^2}$ with $\sigma_\mathrm{\eta}^2 \triangleq  N_0  B$~\cite[P.~233]{gallager2008principles}. Choosing $N_\text{os} \!>\! 2$ gives correlated noise and we thus set $N_\text{os} \! =\!2$ for all remaining discussions.
The noise-free $r^\prime[\tip]$ reads
\begin{align*}
  r^\prime_\tip \triangleq  r^\prime[\tip] %
                 =  \left| \, \sum\nolimits_{m = 0}^{M-1} \combchannel[L, m] \cdot s^\prime[\tip-m] \, \right|^2
                 \numberthis
\end{align*}
where $ s^\prime_\tip \!=\! s^\prime[\tip] \!\triangleq\! s(\tip \Tsym^\prime)$ by~\eqref{eq:s_t_cont}, i.e., the $N_\text{os}$-times upsampled version of the sequence $\left\lbrace\ldots,s_{\ti-1}, s_{\ti},\ldots\right\rbrace$, 
and $\combchannel[L,\tip] \!\triangleq\! \combchannel(L,\tip \Tsym^\prime)$ is the sampled,  length $M$, combined impulse response (CIR) $\combchannel(L,t)$ of the transmitter pulse shaping filter and CD, i.e., $\combchannel(L,t) \;\laplace\; H(L,\omega) \cdot G_\text{tx}(\omega)$.
Henceforth, we omit the argument $L$.
Arranging $\bm{r}^\prime[\tip]  \!=\! \big[r^\prime_\tip,\ldots, r^\prime_{\tip+K-1} \big]^{\T}$ gives
\begin{align}
  \bm{r}^\prime[\tip] = \big|\,  \combchannelmat^\prime \cdot \bm{s}^\prime[\tip] \,\big|^2 \;\; \dimRP{K}
  \label{eq:sampled_r_prime}
\end{align}
with standard linear convolution matrix $\combchannelmat^\prime \! \dimC{K\times N}$, $N \!=\! K\!\,+M\!\,-1$ with (right) shifted versions of the CIR vector $\bm{\combchannel}\!=\!\left[ \combchannel_{M-1}, \ldots, \combchannel_0 \right]^{\T}$ arranged as its rows.
The vector
$\bm{s}^\prime[\tip] \!=\! \big[s^\prime_{\tip - M +1},\ldots, s^\prime_\tip,\ldots,s^\prime_{\tip + K-1}  \big]^{\T} \!\dimRP{N}$ contains the upsampled transmit symbols.
Note that by zero-insertion of upsampling,
\begin{align}
   s^\prime_{k + \Delta} =
   \begin{cases}
     \, s_{\nu}, &\text{if $\left(\tip+\Delta\right)$  is even } \\
     \, 0,                  & \text{otherwise}
   \end{cases}
   \label{eq:even_odd_upsampling}
\end{align}
with $\nu \!=\! \frac{\kappa+\Delta}{2}$ and $\Delta \dimZ{}$, following from $s_{\tip}^\prime \triangleq s(\tip\Tsym^\prime) $ by~\eqref{eq:s_t_cont}.
We thus remove entries with odd index from the vector $\bm{s}^\prime[\tip]$ and denote the result as $\bm{s}[\kappa] \dimRP{ N^\prime}$.
We also discard the according columns of $\combchannelmat^\prime$ to get $ \combchannelmat \! \dimR{K\times N^\prime}$. Note that $\left\lvert \;\uarg\; \right\rvert^2$ is applied element-wise. %
The input-output relationship  reads as
\begin{align}
  \bm{u}[\kappa] = \bm{r}^\prime[\kappa]  + \bm{\upeta}[\kappa] = \big|\,  \combchannelmat \cdot \bm{s}[\tip] \,\big|^2 +  \bm{\upeta}[\tip] \;\dimR{K}
  \label{eq:phase_retrieval_problem}
\end{align}
and $\bm{u}[\kappa]\!=\!  \left[u_\tip,\,\ldots,\, u_{\tip+K-1} \right]^{\T}$ and $\bm{\upeta}[\kappa] \!=\!  \left[\eta_\tip,\,\ldots,\, \eta_{\tip+K-1} \right]^{\T}$. 

%% file: Figures/detailed_system_model.tex
\usetikzlibrary{decorations.markings}
\tikzset{node distance=2.3cm}

\pgfdeclarelayer{background}
\pgfdeclarelayer{foreground}
\pgfsetlayers{background,main,foreground}
\tikzset{boxlines/.style = {draw=black!20!white,}}
\tikzset{boxlinesred/.style = {densely dashed,draw=red!50!white,thick}}

\pgfmathsetmacro{\samplerwidth}{30}

\tikzset{midnodes/.style = {midway,above,text width=1.5cm,align=center,yshift=-1.3em}}
\tikzset{midnodesRP/.style = {midway,above,text width=1.5cm,align=center,yshift=-1.5em}}

\begin{tikzpicture}[]
    \node [input, name=input] {Input};
    \node [comblock,right of=input,node distance=1.5cm,text width=1.2cm,align=center,font=\footnotesize] (geomshaping) {$\sqrt{\left(\uarg\right)}$};
    \node [comblock,right of=geomshaping,node distance=2.1cm] (txfilter) {$g_\text{tx}(t)$};
    \node [comblock,right of=txfilter] (cir) {$h(t,L)$};
    \node [comblock,right of=cir] (sld) {$\left\lvert \,\cdot\, \right\rvert^2$};
    \node [sum,right of=sld,node distance=1.7cm] (sumnode) {$+$};
    \node [input, name=noise,above of=sumnode,node distance=0.7cm] {Input};
    \node [comblock,right of=sumnode,node distance=1.7cm] (rxfilter) {$g_\text{rx}(t)$};
    \node [comblock,right of=rxfilter,minimum width=\samplerwidth pt,node distance=2.1cm] (sampler) {};
    \node [comblock,right of=sampler,node distance=2.7cm] (dsp) {DSP};
    \node [input, name=output, right of=dsp,node distance=1.3cm] {Output};
    \draw[thick] (sampler.west) -- ++(\samplerwidth/4 pt,0) --++(\samplerwidth/2.7 pt,\samplerwidth/4.5 pt );
    \draw[thick] (sampler.east) -- ++(-\samplerwidth/3pt,0);

    \draw ($(sampler.west) + (\samplerwidth/4.5 pt,0.25)$)edge[out=0,in=100,-latex,thick] ($(sampler.east) + (-\samplerwidth/2.5 pt,-0.2)$);

    \draw[-latex,thick] (input) node[midway,above,xshift=0.2cm,align=center,midnodesRP,text width=0.45cm](){$\breve{s}_\ti$\\[0.29em] $\,\mathbb{R}_+$ } -- (geomshaping);
    \draw[-latex,thick] (geomshaping) -- node[midnodesRP](s_ti){$s_\ti$\\[0.29em] $\mathbb{R}_+$ } (txfilter);

    \draw[-latex,thick] (txfilter) -- node[midnodes](){$q(0,t)$\\[0.2em] $\mathbb{R}$} (cir);
    \draw[-latex,thick] (cir) -- node[midnodes](){$q(L,t)$\\[0.2em] $\mathbb{C}$} (sld);
    \draw[-latex,thick] (sld) -- node[midnodes,yshift=-0.2em](){$r^\prime(t)$\\[0.2em] $\mathbb{R}_+$} (sumnode);
    \draw[-latex,thick] (sumnode) -- node[midnodes](){$r(t)$\\[0.2em] $\mathbb{R}$} (rxfilter);
    \draw[-latex,thick] (rxfilter) -- node[midnodes](){$u(t)$\\[0.2em] $\mathbb{R}$} (sampler);
    \draw[-latex,thick] (sampler) --  node[midnodes](){$u(\nu \Tsym^\prime)$\\[0.2em] $\mathbb{R}$} (dsp)  ;
    \draw[-latex,thick] (dsp) --  node[midnodesRP,xshift=0cm,align=center](){$\hat{s}_\ti$\\[0.29em] $\mathbb{R}_+$}(output);

    \draw[-latex,thick] (noise) -- (sumnode);
    \node[above] () at (noise) {$\eta^\prime(t) \in\, \mathbb{R}$};
    \node[above,yshift=1em] () at (sampler) {$ \ti \Tsym^\prime$};

    \node[above,xshift=-0.35cm,yshift=0.6cm,text width=3cm,font=\footnotesize,align=center] () at (geomshaping) {Relevant\\ from Sec.~\ref{sec:tx_symbol_dist}};

    \pgfmathsetmacro{\blockheight}{16}
    \draw[draw=black,densely dashed] (input) ++(0,\blockheight pt) rectangle ($(s_ti) + (-0.29,-\blockheight pt)$);

    \begin{pgfonlayer}{background}
        \draw[boxlinesred] ($( txfilter |- noise) + (0,+20pt)$) -- node[left,yshift=1.05cm,text width=1.2cm,rotate=90,font=\scriptsize  ,align=right,]{Electrical\\[0.2em] Optical}  ($( txfilter |- noise) + (0,-33pt)$);
    \end{pgfonlayer}
    \begin{pgfonlayer}{background}
        \draw[boxlinesred] ($( sld |- noise) + (0,+20pt)$) -- node[left,yshift=1.05cm,text width=1.2cm,rotate=90,font=\scriptsize  ,align=right,]{Optical\\[0.2em]Electrical}  ($(sld |- noise) + (0,-33pt)$);
    \end{pgfonlayer}

\end{tikzpicture}

%% file: includes/4_lmmse.tex
\section{WF Problem Statement}
\label{sec:lmmse}
We now formulate the optimization problem to obtain the Wiener Filter~\cite[P. 382]{kay1993fundamentals}:
\begin{align}
   \min_{\bm{g},\meanlmmse} \, \mathrm{MSE}\text{, } \;\;\mathrm{MSE} = \underset{\bm{s},\bm{\upeta}}{\E} \,\,\left|\, \hat{s}^{\vphantom{\prime}}_{\tip} - s^\prime_{\tip} \,\right|_2^2
  \label{eq:lmmse_estimation_problem}
\end{align}
where the MSE is formulated between a single estimate $ \hat{s}[\tip] \triangleq \hat{s}_{\tip} = \bm{g}^{\T} \bm{u}[\tip] + \meanlmmse$ with filter vector $\bm{g} \dimR{K\times 1}$, filter mean $\meanlmmse \dimR{}$ and a single transmitted symbol
$s^\prime_{\tip}$.
Hence, $s^\prime_{\tip}$ is \text{linearly} estimated from a vector of $K$ measurements $\bm{u}[k]$.
By~\eqref{eq:even_odd_upsampling}, optimization~\eqref{eq:lmmse_estimation_problem} needs to \textit{only} be solved for even $\kappa$. 
\subsection{WF Estimate}
Letting $\tip$ be \textit{even}, and omitting $\tip$ for vectors, the solution of~\eqref{eq:lmmse_estimation_problem} reads~\cite[P.$\;$382]{kay1993fundamentals},
\begin{align}
   \hat{s}_\tip \!=\! \left(\covm{c}{\mathrm{s}^\prime_\tip \mathbf{u}}  \covm{C}{\mathbf{u}\mathbf{u}}^{-1}\right) \! \cdot \bm{u}  \;+ \left(\mu_\mathrm{s}  - \covm{c}{\mathrm{s}^\prime_\tip \mathbf{u}} \covm{C}{\mathbf{u}\mathbf{u}}^{-1} \cdot  \bm{\mu}_{\mathbf{u}} \right)
   \triangleq  \lmmse^{\T} \bm{u} + \meanlmmse
  \label{eq:lmmse_solution}
\end{align}
where the covariance matrices and mean vector compute as 
\newcommand{\wvar}{\ensuremath{\bm{w}}} %
\newcommand{\zvar}{\ensuremath{\bm{z}}} %
\pgfmathsetmacro{\linebreakspacing}{-3}
\begingroup
\allowdisplaybreaks
\begin{align*}
  \covm{c}{\mathrm{s}^\prime_\tip\mathbf{u}} &= 2 \sigma_\mathrm{s}^2 \mu_\mathrm{s} \cdot \Real{ \left( \combchannelmat  \cdot \unitvec{  M^\prime }\right)  \circ \wvar^{*}  }^{\T}   %
  \label{eq:covm_csu}
  \numberthis
  \\[-3 pt]
  \mean{\mathbf{u}} &=    \sigma_\mathrm{s}^2 \diag\left(  \combchannelmat\combchannelmat^{\h}  \right)   + \mu_\mathrm{s}^2  \big|\wvar \big|^{\circ 2}   %
  \numberthis
  \label{eq:mu_u} \\[-3 pt]
  \covm{C}{\mathbf{u}\mathbf{u}} &=  
  \left(\moment{4} - 3 \sigma_\text{s}^4 \right)  \cdot  \big| \combchannelmat \big|^{\circ 2}  \cdot \big| \combchannelmat \big|^{\circ 2,\T} \\
  &+  \sigma_\mathrm{s}^4 \cdot 
   \left( \zvar \zvar^{\T} + \big| \combchannelmat \combchannelmat^{\h} \big|^{\circ 2} +  \big| \combchannelmat \combchannelmat^{\T} \big|^{\circ 2} \right) \\[\linebreakspacing pt]
  & + \sigma_\mathrm{s}^2 \mu_\mathrm{s}^2 \cdot  \bigg[ \zvar \cdot {\big| \wvar \big|^{\circ 2}}^{\T} +  \big| \wvar \big|^{\circ 2} \cdot \zvar^{\T}  \\[\linebreakspacing pt]
  & + 2 \cdot \Real{ \diag{\left( \wvar^* \right)} \combchannelmat \combchannelmat^{\T} \diag{\left( \wvar^* \right)}  } \\[\linebreakspacing pt]
  & + 2 \cdot \Real{ \diag{\left( \wvar^* \right)} \combchannelmat \combchannelmat^{\h} \diag{\left( \wvar \right)}  }   \bigg] \\[-6 pt]
  & + \mu_\mathrm{s}^4 \cdot \big| \wvar \big|^{\circ 2} \cdot {\big| \wvar \big|^{\circ 2}}^{\T}
    + \sigma_\mathrm{\eta}^2 \cdot \mident{K \times K }   - \mean{\mathbf{u}} \mean{\mathbf{u}}^{\T}
   ,
  \label{eq:covm_cuu}
  \numberthis
\end{align*}
\endgroup
and $\combchannelmat \dimC{K \times N^\prime}$, $N^\prime \! = \! M^\prime \!+  K^\prime \! + 1$ ,
with $M^\prime =  \lfloor \frac{M-1}{N_\text{os}} \rfloor,\;\; K^\prime = \lfloor  \frac{K-1}{N_\text{os}}  \rfloor$ and vectors $  \wvar \!=\! \combchannelmat \!\cdot \mones{N^\prime}$,
$  \zvar \!=\! \big| \combchannelmat \big|^{\circ 2} \cdot \mones{N^\prime}$. For computation of $\covm{C}{\mathbf{u}\mathbf{u}}$ we assumed the transmit symbols $s_\ti$ to be independent and identically distributed (iid) by a symmetric (around the mean) PDF with mean $\mu_\mathrm{s}$ and variance $\sigma_\mathrm{s}^2$ and denote the fourth-order \textit{central} moment of $s_\ti$ by $\moment{4}$. %
All quantities~\eqref{eq:covm_csu}--\eqref{eq:covm_cuu} are real-valued, $\covm{C}{\mathbf{u}\mathbf{u}}$ is positive definite and invertible. 
The noise is iid with $\bm{\upeta}[\tip] \sim \GaussianPDF{\mnull{K}}{\covm{C}{\bm{\upeta}\bm{\upeta}}}$,
where $\covm{C}{\bm{\upeta}\bm{\upeta}} = \sigma_\mathrm{\eta}^2 \!\cdot\! \mident{K \times K }$ and $\sigma_\mathrm{\eta}^2 = N_0B$.
The WF can be pre-computed offline
and efficiently applied using the FFT and overlap-add processing. 
We note that the WF is the optimal \textit{affine estimator} in the MSE sense. However, it is only Bayesian MSE optimal for jointly Gaussian $\bm{u}$ and $s_\nu$, which is not the case here~\cite[P. 382]{kay1993fundamentals} and hence better \textit{nonlinear} estimators will exist.
\subsection{Mismatched WF}
\label{sec:tx_symbol_dist}
The $s_\text{\ti}$ are iid by a symmetric probability mass function (PMF) with support $\mathcal{S}$.
Considering Fig.~\ref{fig:Detailed_Flow_Graph_IM_DD} without CD, zero noise, and Nyquist property of $g_\text{tx}(t)$ gives
\begin{align*}
 u[\tip] \!=\!  \bigg(\sum\limits_{\ti=-\infty}^{\infty} s_\ti  g_\text{tx}(\kappa\Tsym^\prime -\nu\Tsym) \bigg)^2
  \!=\! \begin{cases}
    \,s_{v}^2: \,\text{even}\;\tip\! \;\;  \\
    \,\text{ISI}: \,\text{odd}\;\;\tip\!  \,
  \end{cases}
  \hspace{-10pt}
  \numberthis
  \label{eq:tx_distribution_sampling_wo_dispersion}
\end{align*}
where for even $\tip$, $\ti \!=\! \frac{\tip}{2}$ by~\eqref{eq:even_odd_upsampling}.
The ISI corresponds to $u[\tip]$ with odd  $\tip$
and is discarded in this simple model. At even $\tip$, we get the squared transmitted data.
Choosing $s_\ti \dimRP{ }$
allows to unambiguously recover the transmitted symbols (already incorporated in Figs.~\ref{fig:Signal_Flow_Graph_IM_DD},\ref{fig:Detailed_Flow_Graph_IM_DD}). In addition, the receive symbols PMF has support $\mathcal{S}^\prime \!=\! \left\{ s^2\,|\, s\in \mathcal{S} \right\}$, which makes it 
involved
for a \textit{linear} WF to map $\mathcal{S}^\prime$ onto $\mathcal{S}$. With AWGN, we also notice that symbols with smaller amplitudes are more affected by noise, which is undesired.
From now on we include a pre-distortion block (cf. Fig.~\ref{fig:Detailed_Flow_Graph_IM_DD}) in our discussions, which yields
\begin{align}
  s_\ti = \sqrt{\breve{s}_\ti}, \quad \text{ with }\quad
  \breve{s}_\ti \in \mathcal{S},\;\; s_\ti \in  \sqrt{\mathcal{S}}
    \label{eq:pre_distortion_sqrt}
\end{align}
where $\breve{s}_\ti$, $s_\ti$  have mean $\breve{\mu}_\mathrm{s}$, $\mu_\mathrm{s}$ and variance ${\breve{\sigma}_\mathrm{s}^{2}}$,  ${\sigma_\mathrm{s}^{2}}$, respectively, leading to $s_\ti^2 \in \mathcal{S}$. Note that~\eqref{eq:pre_distortion_sqrt} is applied on the electrical side, i.e.,  in the transmitter DSP (cf. Fig.~\ref{fig:Detailed_Flow_Graph_IM_DD}).
Considering~\eqref{eq:pre_distortion_sqrt}, the WF needs to be recomputed based on%
\begin{equation}
  \mathrm{MSE}^\prime = \underset{\bm{s},\bm{\upeta}}{\E} \,\,\left|\, \hat{s}_\tip - s^\prime_\tip \;\right|_2^2 %
\label{eq:mse_prime}
\end{equation}
where now in analogy to~\eqref{eq:even_odd_upsampling}, for even $\tip$ and $\nu \!=\! \frac{\tip}{2}$, we get $s_{\tip}^\prime \!=\! \breve{s}_\ti \in \mathcal{S}$  and
$\sqrt{s^{\prime}_{\tip}} \!=\! \sqrt{\breve{s}_\ti} \in \! \mathcal{\sqrt{S}}$. Note that the predistortion is incorporated in $\hat{s}_\tip$. %
To facilitate analytic expressions, we calculate a \textit{mismatched} WF 
by approximating $\sqrt{\left( \mathbf{\cdot}\right)}$ with a first-order Taylor series around the mean $\breve{\mu}_\mathrm{s}$ of $\breve{s}_\ti$,
\begin{align}
  \sqrt{ \breve{s}_\ti} \approx %
   t_\alpha\cdot  \breve{s}_\ti + t_\beta
  \label{eq:taylor_series_first_order}
\end{align}
with constants $t_\alpha \!=\! 1/(2 \sqrt{\breve{\mu}_\mathrm{s}}),\,t_\beta  \!=\! \sqrt{\breve{\mu}_\mathrm{s}}/2$.
We find the mismatched WF by substitutions for \textit{all} quantities ~\eqref{eq:covm_csu}--\eqref{eq:covm_cuu}:
\begin{align*}
  t_\alpha \breve{\mu}_\mathrm{s} + t_\beta \longrightarrow \mu_\mathrm{s}; \;\;\;\;\;
  t_\alpha^2 {\breve{\sigma}_\mathrm{s}^2}  \longrightarrow \sigma_\mathrm{s}^2;\;\;\;\;\;
  t_\alpha^4  \breve{\mu}_4  \longrightarrow \moment{4}
  \numberthis
\end{align*}
where $\breve{\mu}_{4}$ is the fourth-order central  moment of $\breve{s}_\ti$.  We obtain the modified ${\bm{g}^{\T}}\!$ and $\meanlmmse$ as a function of $\breve{\mu}_\mathrm{s}$, $\breve{\sigma}_\mathrm{s}^2$ and $\breve{\mu}_4$,
\begin{align*}
   {\bm{g}^{\T}} = t_\alpha^{-1} \cdot \covm{c}{\mathrm{s}^\prime_\tip\mathbf{u}}  \covm{C}{\mathbf{u}\mathbf{u}}^{-1}, \quad\quad
   \meanlmmse =  \breve{\mu}_\mathrm{s} -  t_\alpha^{-1}  \cdot \covm{c}{\mathrm{s}^\prime_\tip \mathbf{u}} \covm{C}{\mathbf{u}\mathbf{u}}^{-1} \cdot  \mean{\mathbf{u}}
   \numberthis
   .
\end{align*}
\subsection{SNR Definition}
Since noise is added on the electrical side, we define the SNR in the electrical domain~\cite[P. 153]{AgrawalFourthEdFiberOptics} and get the average electrical receive power as
\begin{align*}
	P_\text{rx,el} %
	\!=\! \frac{\sum\nolimits_{\tip = 0}^{K-1} \E_s   \big| r^\prime(\tip \Tsym^\prime) \big|^2}{N^\prime  N_\text{os}}
	\!= \! \frac{\trace{\left( \covm{C}{\mathbf{r}^\prime \mathbf{r}^\prime}\right) + \big|\big| \mean{\mathbf{r}^\prime}  \big|\big|_2^2}}{N^\prime N_\text{os} }
	\numberthis
  \label{eq:receiver_snr_el_covm}
\end{align*}
with $\covm{C}{\mathbf{r}^\prime \mathbf{r}^\prime} =
\covm{C}{\mathbf{u}\mathbf{u}} - \covm{C}{\bm{\upeta}\bm{\upeta}}  \dimR{K \times K}$, $\mean{\mathbf{r}^\prime} = \mean{\mathbf{u}} \dimR{K}$,
where~\eqref{eq:covm_csu}-\eqref{eq:covm_cuu} are computed numerically.
The electrical SNR after sampling is then given as
\begin{align}
	\text{SNR}_\text{el} = P_\text{rx,el}/ \sigma_\mathrm{\upeta}^2.
\end{align}
For constant $P_{\textrm{tx,opt}} \!=\!\sigma_\mathrm{s}^2 \!\,+\,\! \mu_\mathrm{s}^2$, we remark that by~\eqref{eq:mu_u}-\eqref{eq:covm_cuu}, $\text{SNR}_\text{el}$ 
depends on the particular choice of $\sigma_\mathrm{s}^2$, $\mu_\mathrm{s}^2$
\subsection{Geometric Shaping}
\label{sec:geom_shaping}
The achievable rate for different SNRs depends on the constellation mean  $\mu_\mathrm{s}$ and its variance  $\sigma_\mathrm{s}^2$, with $\sigma_\mathrm{s}^2 + \mu_\mathrm{s}^2 = P_\text{tx,opt}$. In the following, we let $\breve{s}_\nu \in\mathcal{S}$, where
\begin{align}
  \mathcal{S}=\bigg\{P_{\textrm{tx,opt}}-\frac{D}{2}+\frac{i}{Q-1}D\bigg\}_{i=0}^{Q-1}
  \label{eq:tx_distribution_wo_sqrt}
\end{align}
has equal-distance spaced elements and $D\le 2P_{\textrm{tx,opt}}$ denotes the \textit{constellation span}.
Though not equivalent to rate, we use the \textit{error to signal power ratio} (ESR),
\begin{align}
  \text{ESR}= \frac{\mathrm{MSE}^\prime}{\breve{\sigma}_\mathrm{s}^2} =
  1- \frac{\covm{c}{\mathrm{s}^\prime_\tip\mathbf{u}} \cdot \covm{C}{\mathbf{u}\mathbf{u}}^{-1}\cdot  \covm{c}{\mathrm{s}^\prime_\tip\mathbf{u}}^{\T}}{\breve{\sigma}_\mathrm{s}^2}
  \label{eq:esr}
\end{align}
as a proxy to find good values for $\sigma_\mathrm{s}^2$ and  $\mu_\mathrm{s}$ for a particular SNR. The ESR expression in closed form enables the solution of a corresponding optimization problem of low computational complexity.
Varying $D$ for a fixed $P_{\textrm{tx,opt}}$, we obtain constellations $\sqrt{\mathcal{S}}$ with varying spacing and distance from zero (cf. Fig.~\ref{fig:achievable_rate_esr_geom_shaping} (b)), 
thus varying $\mu_\mathrm{s}$ and $\sigma_\mathrm{s}^2$. 
At low SNR, larger constellation spacing helps mitigate the effects of AWGN; at high SNR, constellations further away from zero lead to fewer SLD ambiguities. %

%% file: includes/5_numerical_results.tex
\section{Numerical Results}
\label{sec:numerical_results}
\begin{figure}[tbp]\hspace*{11pt}
  \begin{subfigure}{1.1\columnwidth}
  \input{Figures/0_achievable_rates.tex}
  \end{subfigure}
    \hspace*{0pt}
    \begin{subfigure}{1\columnwidth} %
      \vspace{1pt}
      \input{Figures/0_constellations_vs_snr.tikz}%
    \end{subfigure} %
  \hspace*{3pt}
  \begin{subfigure}{.44\columnwidth}
    \centering
    \input{Figures/0_geom_shaping.tex}%
  \end{subfigure}%
  \hspace*{15pt}
  \begin{subfigure}{.44\columnwidth}
    \centering
     \input{Figures/0_esr.tex}%
  \end{subfigure} 
  \caption{
\textbf{(a)} Achievable rates, \textbf{(b)} optimized $8$-PAM constellation $\sqrt{\mathcal{S}}$, \textbf{(c)} ESR-optimal constellation span, \textbf{(d)} ESR, versus SNR in dB. %
Empty and filled markers denote $L_\text{btb}\!=\!\SI{0}{\kilo\meter}$ (plots a,$\,$d) and $L\!=\!\SI{20}{\kilo\meter}$ (plots a,$\,$b,$\,$c$\,$d), respectively.
}
\label{fig:achievable_rate_esr_geom_shaping}
\vspace{-11pt}
\end{figure}
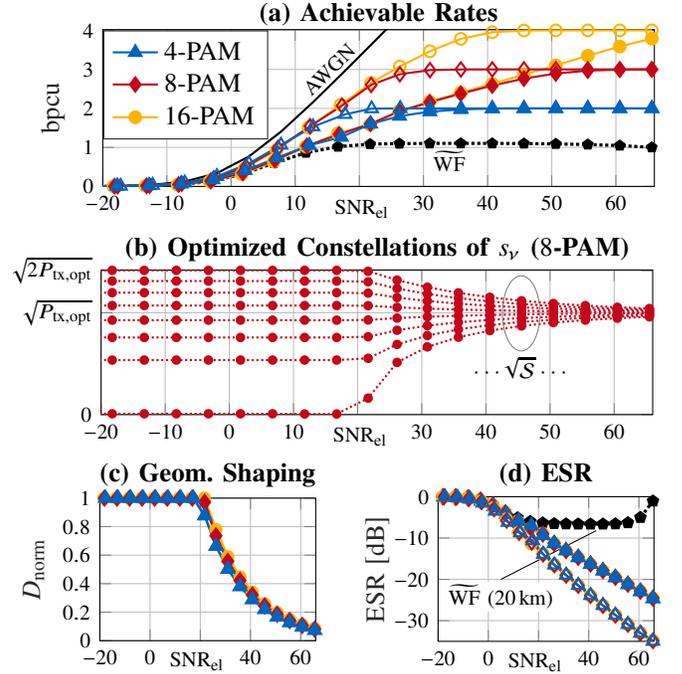

We consider a standard single-mode fiber (SSMF) at wavelength $\SI{1550}{nm}$, with $\beta_2 \!=\! \SI{-2.168e-23}{\second^2\per\kilo\meter}$, $\alpha  \!=\!  \SI{0.046}{\per\kilo\meter}$, $\gamma  \!=\!  \SI{1.27}{\per\watt\per\kilo\meter}$,
link length $L\!=\!\SI{20}{\kilo\meter}$,
receiver oversampling $N_\text{os} \!=\! 2$, symbol rate $B \!=\! f_\text{sym} \!=\! \SI{27}{\giga Baud}$ and modulation alphabets \{$4$,$8$,$16$\} unipolar PAM. 
We transmit \SI{100e3}{} symbols and %
set %
$P_\text{tx,opt} \!=\! \phi_{\text{max}}/(\gamma L_\text{eff} )$, with maximal phase rotation
$\phi_{\text{max}}  \!=\! 0.1$, $L_\text{eff} \!=\! \left[1\!-\!\exp{(-\alpha L)}\right]/ \alpha$~\cite[P.  64]{AgrawalFourthEdFiberOptics}, where the Kerr nonlinearity becomes negligible and all simulations are then carried out with $\gamma \!=\! 0$.
The transmitter uses ESR-optimal geometric shaping and the receiver applies the \WF.
We choose the achievable rate from~\cite{mi_computation_javier_garcia} as the performance metric, which is a lower bound on the mutual information of the channel.
Comparisons are made to the capacity $\frac{1}{2} \cdot \log_2(1 + \mathrm{SNR}_\text{el})$ of the real-valued AWGN channel,
a CD-free back-to-back scenario ($L_\text{btb}\!=\!\SI{0}{\kilo\meter}$) and a naive Wiener Filter \WFlin, which discards \textit{all} non-linearities in Fig.~\ref{fig:Detailed_Flow_Graph_IM_DD}. %
 The naive \WFlin~computes in analogy to~\eqref{eq:lmmse_solution} with
\begin{align*}
  \covm{\mathbf{C}}{\mathbf{u}\mathbf{u}} &= \breve{\sigma}_\mathrm{s}^2  \combchannelmat \combchannelmat^{\h} + \covm{\mathbf{C}}{\bm{\upeta}\bm{\upeta}},\;\;\;
  \covm{\mathbf{c}}{\mathrm{s}_\tip^\prime \mathbf{u}} \!=\! \breve{\sigma}_\mathrm{s}^2  \left(\combchannelmat \unitvec{M^\prime} \right)^{\h},\;\;\;
  \mean{\mathbf{u}} = \breve{\mu}_\mathrm{s} \combchannelmat \!\cdot\! \mones{N^\prime}
\end{align*}
and we keep only the real part of its estimate. 
For both \WF s, we set the number of considered observations to the length of the CIR, i.e., $K\!=\!M$.
The sampled CIR $\combchannel(L, \tip \Tsym^\prime)$ is truncated and only elements larger than $\frac{1}{100} \cdot \max_\tip\left\{  \big|\combchannel(L, \tip \Tsym^\prime) \big| \right\}$ are kept in the vector $\bm{\combchannel}$. %
Fig.~\ref{fig:achievable_rate_esr_geom_shaping}~(a) shows achievable rates in bits per channel use (bpcu) for $L_\text{btb} \!=\! \SI{0}{\kilo\meter}$ and $L \!=\! \SI{20}{\kilo\meter}$ plotted against the SNR, when using geometric shaping with optimal normalized constellation span
$D_\text{norm} \!=\! D/(2 P_\text{tx,opt})$ as given in Fig.~\ref{fig:achievable_rate_esr_geom_shaping}~(c). Related constellations $\sqrt{\mathcal{S}}$ for $L\!=\!\SI{20}{\kilo\meter}$ are shown in Fig.~\ref{fig:achievable_rate_esr_geom_shaping}~(b).
For moderate fiber lengths $L$, the \WF~achieves the maximum mutual information with finite input alphabets asymptotically in $\mathrm{SNR}_\text{el}$, as it is able to compensate CD and SLD-nonlinearity in the high-SNR regime. In addition, the \WF~significantly outperforms the suboptimal \WFlin, which saturates to the same rate for all used modulation formats. 
The achievable rate for the \WF~and $L_\text{btb}\!=\!\SI{0}{\kilo\meter}$ has a smaller slope compared to the AWGN channel, which is caused by the ISI of the pulse shaping filter (cf.~\eqref{eq:tx_distribution_sampling_wo_dispersion}) and $N_\text{os} \!=\! 2$. Furthermore, for $L \!=\! \SI{20}{\kilo\meter}$, the slope decreases further, as the derived \WF~is only Bayesian MSE optimal for jointly Gaussian observations $\bm{u}$ and transmit symbols $\breve{s}_\ti$ (cf. Sec. IV-A).
We also observe that the optimal constellation span in Fig.~\ref{fig:achievable_rate_esr_geom_shaping}~(c) decreases in the SNR, as predicted in Sec.~\ref{sec:geom_shaping}.
For verification, Fig.~\ref{fig:achievable_rate_esr_geom_shaping}~(d) shows the empirical ESR, where the $\text{MSE}^\prime$~\eqref{eq:esr} is calculated through simulations. %
Applying \WF~with geometric shaping results in a monotonically decreasing ESR, whereas the ESR for \WFlin~exhibits an error floor. %
For reproducible results, all simulations are available on~\cite{github_repository}.
\newlength{\figurewidth}
\newlength{\figureheight}
\setlength\figurewidth{0.90\columnwidth}%
\setlength\figureheight{0.34\figurewidth}

%% file: Figures/0_achievable_rates.tex
\pgfplotsset{AWGN/.style = {color=black, mark=none}}

\pgfplotsset{PAM4_btb/.style = {color=mycolor1, solid, mark=triangle, mark options={solid,  scale=1.35, fill=white,}}}
\pgfplotsset{PAM8_btb/.style = {color=mycolor2, solid, mark=diamond, mark options={solid, scale=1.35,mycolor2}}}
\pgfplotsset{PAM16_btb/.style = {color=mycolor3, solid, mark=o, mark options={solid,  mycolor3,scale=1.1}}}

\pgfplotsset{PAM4_20km/.style = {color=mycolor1, mark=triangle*, mark options={solid, mycolor1,scale=1.35} }}
\pgfplotsset{PAM8_20km/.style = {color=mycolor2, mark=diamond*, mark options={solid, mycolor2,scale=1.35}}}
\pgfplotsset{PAM16_20km/.style = {color=mycolor3, mark=*, mark options={solid, mycolor3,scale=1.1}}}

\pgfplotsset{naive_PAM4_20km/.style = {line width=1.2pt,color=black, densely dotted, mark=pentagon*, mark options={solid, black,scale=0.8} }}
\pgfplotsset{naive_PAM8_20km/.style = {color=mycolor2, dotted, mark=none, mark options={solid, mycolor2}}}
\pgfplotsset{naive_PAM16_20km/.style = {color=mycolor3, dotted, mark=none, mark options={solid, mycolor3,scale=0.7}}}

\pgfplotsset{
  every axis plot/.append style={line width=0.8pt},
  }

\begin{tikzpicture}

\begin{axis}[%
  width=0.92\figurewidth,
  height=0.77\figureheight,
at={(0.909in,0.494in)},
scale only axis,
xmin=-20,
xmax=66,
xlabel style={font=\color{white!15!black}},
xlabel={$\mathrm{SNR}_\mathrm{el}$},
ymin=0,
ymax=4.02,
ylabel style={font=\color{white!15!black}},
ylabel={bpcu},
axis background/.style={fill=white},
title style={font=\bfseries},
title={(a) Achievable Rates},
xmajorgrids,
ymajorgrids,
legend style={legend cell align=left, align=left, draw=white!15!black},
legend style={at={(0.001,0.995)},anchor=north west},
reverse legend,
title style={yshift=-1.75ex},
xlabel style={font=\footnotesize,at={(0.47,-0.3ex)}},
xtick={-20,-10,0,10,20,30,40,50,60,70},
xticklabels={$-20$,$-10$,$0$,$10$,,$30$,$40$,$50$,$60$},
ticklabel style = {font=\footnotesize}
]

\addplot[naive_PAM4_20km]
  table[row sep=crcr]{%
  -43.3261079685733	0.000412853255126322\\
  -38.3261079685733	0.00044064912050672\\
  -33.3261079685733	0.000532996558616006\\
  -28.3261079685733	0.000832831306551185\\
  -23.3261079685733	0.00179380197263939\\
  -18.3261079685733	0.00484466750759047\\
  -13.3261079685733	0.0144074117857773\\
  -8.32610796857331	0.0434906483536825\\
  -3.32610796857331	0.124927521982515\\
  1.67389203142669	0.310631600583496\\
  6.6738920314267	0.597284723594753\\
  11.6738920314267	0.864092971073592\\
  16.6739159311484	1.02024493467225\\
  21.6740770375788	1.08551143201479\\
  26.237652855904	1.09691893209631\\
  30.9667822044597	1.1028473427634\\
  35.8083129896253	1.10464258885112\\
  40.7163180472178	1.10322208893215\\
  45.6631852884082	1.09863611509046\\
  50.6324322105946	1.08982271118305\\
  55.6145896514778	1.07424237900696\\
  60.6042399730246	1.04734565581339\\
  65.5982781246862	1.00168301610473\\
  };

\addplot[AWGN]
  table[row sep=crcr]{%
  -42.2171942434636	4.32924422761819e-05\\
  -37.2171942434636	0.000136893840980857\\
  -32.2171942434636	0.000432807546286161\\
  -27.2171942434636	0.0013677707557389\\
  -22.2171942434636	0.00431643390224949\\
  -17.2171942434636	0.013562384037358\\
  -12.2171942434636	0.0420442928545435\\
  -7.21719424346356	0.125355399771652\\
  -2.21719424346356	0.33911651178637\\
  2.78280575653644	0.767511792234799\\
  7.78280575653644	1.40386158319381\\
  12.7844644153875	2.1604793843088\\
  17.7846994001685	2.96588932636125\\
  21.925145950758	3.64630363889845\\
  26.3716416394041	4.38189628623237\\
  31.0417453445943	5.15648958102142\\
  35.848078041246	5.95442450013334\\
  40.7363661383163	6.76622484002146\\
  45.6726650848425	7.58608500332824\\
  50.6366133308262	8.4105656525847\\
  55.616299891852	9.23766943653184\\
  60.6049096925612	10.0662582373488\\
  65.5985282329856	10.8956798947638\\
  };

\addplot [PAM16_20km]
  table[row sep=crcr]{%
  -43.3261079685733	0.000412860768997803\\
  -38.3261079685733	0.000440655517484245\\
  -33.3261079685733	0.000533000159513719\\
  -28.3261079685733	0.000832827322865626\\
  -23.3261079685733	0.00179377725315071\\
  -18.3261079685733	0.00484460756714959\\
  -13.3261079685733	0.014407795366737\\
  -8.32610796857331	0.0435045392987067\\
  -3.32610796857331	0.125238648298997\\
  1.67389203142669	0.315490340973907\\
  6.6738920314267	0.636907739720151\\
  11.6738920314267	1.01932167644989\\
  16.6739159311484	1.35940794451981\\
  21.6740770375788	1.58237079774028\\
  26.237652855904	1.93171592366848\\
  30.9667822044597	2.17278062882561\\
  35.8083129896253	2.40186437472357\\
  40.7163180472178	2.63046112985387\\
  45.6631852884082	2.86385316381169\\
  50.6324322105946	3.10423904624435\\
  55.6145896514778	3.34961369460714\\
  60.6042399730246	3.5873588400461\\
  65.5982781246862	3.78898279164462\\
  };

\addplot[PAM16_btb,forget plot]
  table[row sep=crcr]{%
  -42.9056684128607	0.000403708728166308\\
  -37.9056684128607	0.00044783719628505\\
  -32.9056684128607	0.000588785490023369\\
  -27.9056684128607	0.00103683473189378\\
  -22.9056684128607	0.00245623291666632\\
  -17.9056684128607	0.00693338145477185\\
  -12.9056684128607	0.0209156689272504\\
  -7.90566841286067	0.0633647119301223\\
  -2.90566841286068	0.182442420823783\\
  2.09433158713932	0.45925919840343\\
  7.09433158713933	0.928779268109603\\
  12.0945953924374	1.51009045449994\\
  17.094551984994	2.09299434734382\\
  21.8278970529988	2.65102019429884\\
  26.3069934066747	3.07404646463462\\
  31.0026486987015	3.4651230296527\\
  35.8251860047366	3.78237336575105\\
  40.7231501247374	3.95152938710819\\
  45.6651111362329	3.99606640859341\\
  50.6323116369013	3.99993347840632\\
  55.6138632114878	3.99999999987023\\
  60.6035245183553	4.00000000000005\\
  65.5977440379636	3.99999999999999\\
  };

\addplot[PAM8_20km]
  table[row sep=crcr]{%
  -43.1654852396304	0.000235330942730982\\
  -38.1654852396304	0.000268476900036418\\
  -33.1654852396304	0.000376839784964034\\
  -28.1654852396304	0.000725676111394336\\
  -23.1654852396304	0.00183830935426457\\
  -18.1654852396304	0.00535928752551058\\
  -13.1654852396304	0.0163571138546096\\
  -8.16548523963044	0.0495813259634598\\
  -3.16548523963044	0.141294880095587\\
  1.83451476036956	0.347058784756642\\
  6.83451476036956	0.675840714014603\\
  11.8353201635996	1.04110688681175\\
  16.8353438159748	1.33456585474954\\
  21.7292769565594	1.60761180065601\\
  26.2434682638463	1.92349183223519\\
  30.9702156883719	2.15821545537333\\
  35.8102272438587	2.37805812990542\\
  40.7174152753633	2.584571901947\\
  45.6638042939528	2.76350312940925\\
  50.6327721972275	2.89466818626205\\
  55.6147619131688	2.96793539955462\\
  60.6043254036177	2.9947854468209\\
  65.5983124535082	2.99974829872755\\
  };
\addplot[PAM8_btb,forget plot]
  table[row sep=crcr]{%
  -42.6912853393617	0.000215448703679755\\
  -37.6912853393617	0.000267828267082493\\
  -32.6912853393617	0.000433028258260038\\
  -27.6912853393617	0.000954435082985938\\
  -22.6912853393617	0.00259930183819179\\
  -17.6912853393617	0.00777221275177311\\
  -12.6912853393617	0.0238699105432122\\
  -7.69128533936173	0.0723730924990361\\
  -2.69128533936173	0.205900145326038\\
  2.30871466063827	0.504288759303703\\
  7.30871466063827	0.982952374644875\\
  12.3089137298089	1.54201366335168\\
  17.3090229009613	2.07128705747127\\
  21.842000271276	2.60168287189593\\
  26.3170860125427	2.87638224385047\\
  31.0086726516025	2.98143664093288\\
  35.8286482452229	2.99921479717287\\
  40.7251082847058	2.99999621871477\\
  45.6661864051936	3\\
  50.6328918189286	2.99999999999997\\
  55.6141725312246	3.00000000000001\\
  60.6036856816052	3.00000000000004\\
  65.5978224060813	3.00000000000001\\
  };

\addplot[PAM4_20km]
  table[row sep=crcr]{%
  -42.7991973564099	0.000114467096731374\\
  -37.7991973564099	0.000156023720175824\\
  -32.7991973564099	0.000294482293445952\\
  -27.7991973564099	0.000744606346064458\\
  -22.7991973564099	0.00218736069066914\\
  -17.7991973564099	0.00675926240587437\\
  -12.7991973564099	0.0209920356920762\\
  -7.7991973564099	0.0633983735811734\\
  -2.7991973564099	0.17605538181113\\
  2.2008026435901	0.409225494312782\\
  7.2008026435901	0.739215881702664\\
  12.2022216878795	1.05386807063075\\
  17.201772633258	1.26450104600001\\
  21.7650127155081	1.60238081731368\\
  26.2756456226393	1.80313220813193\\
  30.9895098441702	1.91687674806708\\
  35.8215808749927	1.97394868568301\\
  40.7241882530319	1.99478473538942\\
  45.6679077888073	1.99944257886186\\
  50.6352683925081	1.99997375973892\\
  55.616271463451	1.99999976813677\\
  60.605219665062	1.99999999999999\\
  65.5988433086728	2\\
  };

\addplot[PAM4_btb,forget plot]
  table[row sep=crcr]{%
  -42.2171942434636	0.00011173004070173\\
  -37.2171942434636	0.000177753327704622\\
  -32.2171942434636	0.000389430895531007\\
  -27.2171942434636	0.00106357169053446\\
  -22.2171942434636	0.00320000755722099\\
  -17.2171942434636	0.00992631198240979\\
  -12.2171942434636	0.0307779997329508\\
  -7.21719424346356	0.0926624456181521\\
  -2.21719424346356	0.256064408741847\\
  2.78280575653644	0.590210290480713\\
  7.78280575653644	1.06154145722804\\
  12.7844644153875	1.5321344098889\\
  17.7846994001685	1.84871375670466\\
  21.925145950758	1.99424694692094\\
  26.3716416394041	1.99992317520349\\
  31.0417453445943	1.99999999995611\\
  35.848078041246	1.99999999999997\\
  40.7363661383163	2\\
  45.6726650848425	2\\
  50.6366133308262	1.99999999999998\\
  55.616299891852	2.00000000000003\\
  60.6049096925612	1.99999999999997\\
  65.5985282329856	2.00000000000004\\
  };

\legend{$16$-PAM,$8$-PAM, $4$-PAM}

\node[rotate=45,fill=white,font=\footnotesize,inner sep=1pt,outer sep=1pt] at (axis cs:15.3,3) {AWGN};
\node[rotate=0,fill=white,font=\footnotesize,inner sep=1pt,outer sep=1pt] at (axis cs:34,0.65) {\WFlin};

\end{axis}
\end{tikzpicture}%

%% file: Figures/0_constellations_vs_snr.tikz
%
%
\begin{tikzpicture}
\pgfplotsset{const_points/.style = {color=mycolor2,densely dotted,line width=0.2pt,  thick, mark=*, mark options={solid, line width=1.2pt, scale=0.6, fill=mycolor2}}}

\begin{axis}[%
width=0.92\figurewidth,
height=0.71\figureheight,
at={(0.0in,0.494in)},
scale only axis,
xmin=-20,
xmax=66,
ymin=0,
ymax=sqrt(2*0.006)+0.0005,
axis background/.style={fill=white},
scaled y ticks=false,
ytick={0,0.07746,0.1095
},
yticklabels={0,$\sqrt{P_\text{tx,opt}}$, $\sqrt{2 P_\text{tx,opt}}$
},
xmajorgrids,
ymajorgrids,
legend style={legend cell align=left, align=left, draw=white!15!black},
legend style={at={(0.01,0.99)},anchor=north west},
xlabel={$\mathrm{SNR}_\mathrm{el}$},
title style={font=\bfseries},
title={(b) Optimized Constellations of $s_\nu $ ($8$-PAM)},
title style={yshift=-1.4ex},
xlabel style={font=\footnotesize,at={(0.47,-0.3ex)}},
xtick={-20,-10,0,10,20,30,40,50,60,70},
xticklabels={$-20$,$-10$,$0$,$10$,,$30$,$40$,$50$,$60$},
ticklabel style = {font=\footnotesize}
]
\addplot [const_points]
  table[row sep=crcr, each nth point=1,y expr=sqrt(\thisrowno{1}) ]{%
  x y \\
-43.2642838325448	3.98088307918681e-07\\
-38.2642838325448	3.98088307918681e-07\\
-33.2642838325448	3.98088307918681e-07\\
-28.2642838325448	3.98088307918681e-07\\
-23.2642838325448	3.98088307918681e-07\\
-18.2642838325448	3.98088307918681e-07\\
-13.2642838325448	3.98088307918681e-07\\
-8.26428383254482	3.98088307918681e-07\\
-3.26428383254482	3.98088307918681e-07\\
1.73571616745518	3.98088307918681e-07\\
6.73571616745518	3.98088307918681e-07\\
11.7365200842157	2.13743925704668e-07\\
16.7365330968757	2.11201802279709e-07\\
21.6384707407274	0.000156571161720282\\
26.1494085029992	0.00157583039975992\\
30.8726486542135	0.00265204586513229\\
35.7089643831705	0.0034660730450483\\
40.6127415385988	0.00408086346442994\\
45.5563089059166	0.00454593167861676\\
50.5231435986342	0.0048990093923247\\
55.5035273706286	0.0051688489922157\\
60.4919010560813	0.00537541917228997\\
65.4850020486634	0.00553350964927659\\
};

\addplot [const_points]
  table[row sep=crcr, each nth point=1,y expr=sqrt(\thisrowno{1}) ]{%
  x y \\
-43.2642838325448	0.00172082200809783\\
-38.2642838325448	0.00172082200809783\\
-33.2642838325448	0.00172082200809783\\
-28.2642838325448	0.00172082200809783\\
-23.2642838325448	0.00172082200809783\\
-18.2642838325448	0.00172082200809783\\
-13.2642838325448	0.00172082200809783\\
-8.26428383254482	0.00172082200809783\\
-3.26428383254482	0.00172082200809783\\
1.73571616745518	0.00172082200809783\\
6.73571616745518	0.00172082200809783\\
11.7365200842157	0.0017206903335391\\
16.7365330968757	0.00172068851773666\\
21.6384707407274	0.00183237420339237\\
26.1494085029992	0.00284613080199212\\
30.8726486542135	0.00361485613440095\\
35.7089643831705	0.00419630412005524\\
40.6127415385988	0.00463544013389927\\
45.5563089059166	0.00496763171546129\\
50.5231435986342	0.00521983008239553\\
55.5035273706286	0.00541257265374625\\
60.4919010560813	0.00556012278237072\\
65.4850020486634	0.00567304455164688\\
};

\addplot [const_points]
  table[row sep=crcr, each nth point=1,y expr=sqrt(\thisrowno{1}) ]{%
  x y \\
-43.2642838325448	0.00344124592788774\\
-38.2642838325448	0.00344124592788774\\
-33.2642838325448	0.00344124592788774\\
-28.2642838325448	0.00344124592788774\\
-23.2642838325448	0.00344124592788774\\
-18.2642838325448	0.00344124592788774\\
-13.2642838325448	0.00344124592788774\\
-8.26428383254482	0.00344124592788774\\
-3.26428383254482	0.00344124592788774\\
1.73571616745518	0.00344124592788774\\
6.73571616745518	0.00344124592788774\\
11.7365200842157	0.0034411669231525\\
16.7365330968757	0.00344116583367104\\
21.6384707407274	0.00350817724506447\\
26.1494085029992	0.00411643120422431\\
30.8726486542135	0.00457766640366961\\
35.7089643831705	0.00492653519506219\\
40.6127415385988	0.0051900168033686\\
45.5563089059166	0.00538933175230581\\
50.5231435986342	0.00554065077246636\\
55.5035273706286	0.00565629631527679\\
60.4919010560813	0.00574482639245147\\
65.4850020486634	0.00581257945401717\\
};

\addplot [const_points]
  table[row sep=crcr, each nth point=1,y expr=sqrt(\thisrowno{1}) ]{%
  x y \\
-43.2642838325448	0.00516166984767765\\
-38.2642838325448	0.00516166984767765\\
-33.2642838325448	0.00516166984767765\\
-28.2642838325448	0.00516166984767765\\
-23.2642838325448	0.00516166984767765\\
-18.2642838325448	0.00516166984767765\\
-13.2642838325448	0.00516166984767765\\
-8.26428383254482	0.00516166984767765\\
-3.26428383254482	0.00516166984767765\\
1.73571616745518	0.00516166984767765\\
6.73571616745518	0.00516166984767765\\
11.7365200842157	0.0051616435127659\\
16.7365330968757	0.00516164314960541\\
21.6384707407274	0.00518398028673656\\
26.1494085029992	0.00538673160645651\\
30.8726486542135	0.00554047667293827\\
35.7089643831705	0.00565676627006913\\
40.6127415385988	0.00574459347283794\\
45.5563089059166	0.00581103178915034\\
50.5231435986342	0.00586147146253719\\
55.5035273706286	0.00590001997680733\\
60.4919010560813	0.00592953000253223\\
65.4850020486634	0.00595211435638746\\
};

\addplot [const_points]
  table[row sep=crcr, each nth point=1,y expr=sqrt(\thisrowno{1}) ]{%
  x y \\
-43.2642838325448	0.00688209376746756\\
-38.2642838325448	0.00688209376746756\\
-33.2642838325448	0.00688209376746756\\
-28.2642838325448	0.00688209376746756\\
-23.2642838325448	0.00688209376746756\\
-18.2642838325448	0.00688209376746756\\
-13.2642838325448	0.00688209376746756\\
-8.26428383254482	0.00688209376746756\\
-3.26428383254482	0.00688209376746756\\
1.73571616745518	0.00688209376746756\\
6.73571616745518	0.00688209376746756\\
11.7365200842157	0.0068821201023793\\
16.7365330968757	0.00688212046553979\\
21.6384707407274	0.00685978332840865\\
26.1494085029992	0.0066570320086887\\
30.8726486542135	0.00650328694220693\\
35.7089643831705	0.00638699734507608\\
40.6127415385988	0.00629917014230727\\
45.5563089059166	0.00623273182599487\\
50.5231435986342	0.00618229215260802\\
55.5035273706286	0.00614374363833787\\
60.4919010560813	0.00611423361261298\\
65.4850020486634	0.00609164925875775\\
};

\addplot [const_points]
  table[row sep=crcr, each nth point=1,y expr=sqrt(\thisrowno{1}) ]{%
  x y \\
-43.2642838325448	0.00860251768725747\\
-38.2642838325448	0.00860251768725747\\
-33.2642838325448	0.00860251768725747\\
-28.2642838325448	0.00860251768725747\\
-23.2642838325448	0.00860251768725747\\
-18.2642838325448	0.00860251768725747\\
-13.2642838325448	0.00860251768725747\\
-8.26428383254482	0.00860251768725747\\
-3.26428383254482	0.00860251768725747\\
1.73571616745518	0.00860251768725747\\
6.73571616745518	0.00860251768725747\\
11.7365200842157	0.0086025966919927\\
16.7365330968757	0.00860259778147417\\
21.6384707407274	0.00853558637008074\\
26.1494085029992	0.00792733241092089\\
30.8726486542135	0.00746609721147559\\
35.7089643831705	0.00711722842008302\\
40.6127415385988	0.0068537468117766\\
45.5563089059166	0.00665443186283939\\
50.5231435986342	0.00650311284267885\\
55.5035273706286	0.00638746729986842\\
60.4919010560813	0.00629893722269373\\
65.4850020486634	0.00623118416112804\\
};

\addplot [const_points]
  table[row sep=crcr, each nth point=1,y expr=sqrt(\thisrowno{1}) ]{%
  x y \\
-43.2642838325448	0.0103229416070474\\
-38.2642838325448	0.0103229416070474\\
-33.2642838325448	0.0103229416070474\\
-28.2642838325448	0.0103229416070474\\
-23.2642838325448	0.0103229416070474\\
-18.2642838325448	0.0103229416070474\\
-13.2642838325448	0.0103229416070474\\
-8.26428383254482	0.0103229416070474\\
-3.26428383254482	0.0103229416070474\\
1.73571616745518	0.0103229416070474\\
6.73571616745518	0.0103229416070474\\
11.7365200842157	0.0103230732816061\\
16.7365330968757	0.0103230750974085\\
21.6384707407274	0.0102113894117528\\
26.1494085029992	0.00919763281315309\\
30.8726486542135	0.00842890748074425\\
35.7089643831705	0.00784745949508996\\
40.6127415385988	0.00740832348124594\\
45.5563089059166	0.00707613189968392\\
50.5231435986342	0.00682393353274967\\
55.5035273706286	0.00663119096139896\\
60.4919010560813	0.00648364083277448\\
65.4850020486634	0.00637071906349833\\
};

\addplot [const_points]
  table[row sep=crcr, each nth point=1,y expr=sqrt(\thisrowno{1}) ]{%
  x y \\
-43.2642838325448	0.0120433655268373\\
-38.2642838325448	0.0120433655268373\\
-33.2642838325448	0.0120433655268373\\
-28.2642838325448	0.0120433655268373\\
-23.2642838325448	0.0120433655268373\\
-18.2642838325448	0.0120433655268373\\
-13.2642838325448	0.0120433655268373\\
-8.26428383254482	0.0120433655268373\\
-3.26428383254482	0.0120433655268373\\
1.73571616745518	0.0120433655268373\\
6.73571616745518	0.0120433655268373\\
11.7365200842157	0.0120435498712195\\
16.7365330968757	0.0120435524133429\\
21.6384707407274	0.0118871924534249\\
26.1494085029992	0.0104679332153853\\
30.8726486542135	0.00939171775001291\\
35.7089643831705	0.00857769057009691\\
40.6127415385988	0.00796290015071527\\
45.5563089059166	0.00749783193652844\\
50.5231435986342	0.0071447542228205\\
55.5035273706286	0.0068749146229295\\
60.4919010560813	0.00666834444285524\\
65.4850020486634	0.00651025396586861\\
};

\draw[solid,gray] (axis cs:45.5,0.077) ellipse (0.25cm and 0.5cm) node[black,below,yshift=-0.45cm,font=\footnotesize]{$\cdots \mathcal{\sqrt{S}}\cdots $};
\end{axis}
\end{tikzpicture}%

%% file: Figures/0_geom_shaping.tex
\pgfplotsset{PAM4_20km/.style = {line width=1pt,color=mycolor1, mark=triangle*, mark options={solid, mycolor1,scale=1.3} }}
\pgfplotsset{PAM8_20km/.style = {line width=1pt,color=mycolor2, mark=diamond*, mark options={solid, mycolor2,scale=1.3}}}
\pgfplotsset{PAM16_20km/.style = {line width=1pt, color=mycolor3, mark=*, mark options={solid, mycolor3,scale=1}}}
\begin{tikzpicture}

\begin{axis}[%
width=0.36\figurewidth,
height=0.70\figureheight,
scale only axis,
xmin=-20,
xmax=66,
xlabel style={font=\color{white!15!black}},
xlabel={$\mathrm{SNR}_\mathrm{el} $},
ymin=0,
ymax=1,
ylabel style={font=\color{white!15!black}},
ylabel={$D_\text{norm}$},
axis background/.style={fill=white},
title style={font=\bfseries},
title={(c) Geom. Shaping},
xmajorgrids,
ymajorgrids,
legend style={legend cell align=left, align=left, draw=white!15!black},
title style={yshift=-1.2ex},xtick={-40,0,60},
xtick={-40,-20,0,20,40,60},
xticklabels={$-40$,$-20$,$0$,,$40$,$60$},
xlabel style={font=\footnotesize,at={(0.45,-0.2ex)}},
ticklabel style = {font=\footnotesize}
]

\addplot [PAM16_20km,each nth point=1]
  table[row sep=crcr]{%
  -43.3261079685733	0.999933893038648\\
  -38.3261079685733	0.999933893038648\\
  -33.3261079685733	0.999933893038648\\
  -28.3261079685733	0.999933893038648\\
  -23.3261079685733	0.999933893038648\\
  -18.3261079685733	0.999933893038648\\
  -13.3261079685733	0.999933893038648\\
  -8.32610796857331	0.999933893038648\\
  -3.32610796857331	0.999933893038648\\
  1.67389203142669	0.999933893038648\\
  6.6738920314267	0.999933893038648\\
  11.6738920314267	0.999933893038648\\
  16.6739159311484	0.999935664132086\\
  21.6740770375788	0.999947049145703\\
  26.237652855904	0.779165801034656\\
  30.9667822044597	0.590065795130869\\
  35.8083129896253	0.447195330324134\\
  40.7163180472178	0.33951242396042\\
  45.6631852884082	0.258314628631331\\
  50.6324322105946	0.196713096501033\\
  55.6145896514778	0.149676933998608\\
  60.6042399730246	0.113582943977409\\
  65.5982781246862	0.0858934477186156\\
  };
\addlegendentry{data6}

\addplot [PAM8_20km,each nth point=1]
  table[row sep=crcr]{%
  -43.1654852396304	0.999933893038648\\
  -38.1654852396304	0.999933893038648\\
  -33.1654852396304	0.999933893038648\\
  -28.1654852396304	0.999933893038648\\
  -23.1654852396304	0.999933893038648\\
  -18.1654852396304	0.999933893038648\\
  -13.1654852396304	0.999933893038648\\
  -8.16548523963044	0.999933893038648\\
  -3.16548523963044	0.999933893038648\\
  1.83451476036956	0.999933893038648\\
  6.83451476036956	0.999933893038648\\
  11.8353201635996	0.999964023600324\\
  16.8353438159748	0.999964778381839\\
  21.7292769565594	0.971417760513595\\
  26.2434682638463	0.735670736435146\\
  30.9702156883719	0.556999001668547\\
  35.8102272438587	0.422058486812444\\
  40.7174152753633	0.320453875125566\\
  45.6638042939528	0.243841054073955\\
  50.6327721972275	0.185712498012974\\
  55.6147619131688	0.141303526614213\\
  60.6043254036177	0.107246519068842\\
  65.5983124535082	0.0811005666997563\\
  };
\addlegendentry{data5}
\addplot [PAM4_20km,each nth point=1]
  table[row sep=crcr]{%
  -42.7991973564099	0.999933893038648\\
  -37.7991973564099	0.999933893038648\\
  -32.7991973564099	0.999933893038648\\
  -27.7991973564099	0.999933893038648\\
  -22.7991973564099	0.999933893038648\\
  -17.7991973564099	0.999933893038648\\
  -12.7991973564099	0.999933893038648\\
  -7.7991973564099	0.999933893038648\\
  -2.7991973564099	0.999933893038648\\
  2.2008026435901	0.999933893038648\\
  7.2008026435901	0.999933893038648\\
  12.2022216878795	0.999965167531168\\
  17.201772633258	0.999956303447302\\
  21.7650127155081	0.876898976681359\\
  26.2756456226393	0.663253411125959\\
  30.9895098441702	0.501655585443739\\
  35.8215808749927	0.379895340551333\\
  40.7241882530319	0.288445978090064\\
  45.6679077888073	0.219589778675867\\
  50.6352683925081	0.167342846650977\\
  55.616271463451	0.12738631029127\\
  60.605219665062	0.0966729256578893\\
  65.5988433086728	0.0730999518953372\\
  };
\addlegendentry{data4}

\legend{}

\end{axis}
\end{tikzpicture}%

%% file: Figures/0_esr.tex
\pgfdeclarelayer{background}
\pgfdeclarelayer{foreground}
\pgfsetlayers{background,main,foreground}   %

\pgfplotsset{PAM4_btb/.style = {color=mycolor1, solid, mark=triangle, mark options={solid,  scale=1.35, fill=white,}}}
\pgfplotsset{PAM8_btb/.style = {color=mycolor2, solid, mark=diamond, mark options={solid, scale=1.35,mycolor2}}}
\pgfplotsset{PAM16_btb/.style = {color=mycolor3, solid, mark=o, mark options={solid,  mycolor3,scale=1.1}}}

\pgfplotsset{PAM4_22km/.style = {line width=1pt,color=mycolor1, solid, mark=triangle*, mark options={solid, scale=1.3,mycolor1} }}
\pgfplotsset{PAM8_22km/.style = {line width=1pt,color=mycolor2,solid, mark=diamond*, mark options={solid, scale=1.3,mycolor2}}}
\pgfplotsset{PAM16_22km/.style = {line width=1pt,color=mycolor3,solid, mark=*, mark options={solid, mycolor3,scale=1}}}

\pgfplotsset{naive_PAM4_22km/.style = {line width=1.2pt,color=black, densely dotted, mark=pentagon*, mark options={solid, black,scale=0.9,} }}
\pgfplotsset{naive_PAM8_22km/.style = {line width=1pt,color=mycolor2, dotted, mark=none, mark options={solid, mycolor2,scale=1.3,}}}
\pgfplotsset{naive_PAM16_22km/.style = {line width=1pt,color=mycolor3, dotted, mark=none, mark options={solid, mycolor3,scale=1}}}

\pgfplotsset{
  every axis plot/.append style={line width=0.8pt},
  }

\begin{tikzpicture}

\begin{axis}[%
width=0.36\figurewidth,
height=0.71\figureheight,
scale only axis,
xmin=-20,
xmax=66,
xlabel style={font=\color{white!15!black}},
xlabel={$\mathrm{SNR}_\mathrm{el}$},
ymin=-35,
ymax=0.01,
ylabel style={font=\color{white!15!black}},
ylabel={ESR $[\SI{}{dB}]$},
axis background/.style={fill=white},
title style={font=\bfseries},
title={(d) ESR},
xmajorgrids,
ymajorgrids,
legend style={legend cell align=left, align=left, draw=white!15!black},
title style={yshift=-1.2ex},
xtick={-40,-20,0,20,40,60},
xticklabels={$-40$,$-20$,$0$,,$40$,$60$},
xlabel style={font=\footnotesize,at={(0.45,-0.2ex)}},
ylabel shift = -5pt,
ticklabel style = {font=\footnotesize}
]
\draw[-] (axis cs:5,-20) node[below,fill=white,font=\footnotesize,inner sep=1pt,outer sep=1pt]{\WFlin~(\SI{20}{\kilo\meter})} --  (axis cs:43,-8.2);

\addplot [naive_PAM4_22km,each nth point=1]
  table[row sep=crcr]{%
  -43.3261079685733	-0.00123197108200768\\
  -38.3261079685733	-0.00139662902082686\\
  -33.3261079685733	-0.00194662332922662\\
  -28.3261079685733	-0.00373732488319615\\
  -23.3261079685733	-0.00948504020852219\\
  -18.3261079685733	-0.0277466565721883\\
  -13.3261079685733	-0.0850007488001404\\
  -8.32610796857331	-0.259052829584239\\
  -3.32610796857331	-0.745520278115268\\
  1.67389203142669	-1.85104558388046\\
  6.6738920314267	-3.55269375171913\\
  11.6738920314267	-5.11303375745159\\
  16.6739159311484	-5.98998963610908\\
  21.6740770375788	-6.34334361036783\\
  26.237652855904	-6.51698185706032\\
  30.9667822044597	-6.57940883380194\\
  35.8083129896253	-6.60997075752271\\
  40.7163180472178	-6.62212307482524\\
  45.6631852884082	-6.6169361553646\\
  50.6324322105946	-6.56650829227695\\
  55.6145896514778	-6.29906339893112\\
  60.6042399730246	-5.01474223364536\\
  65.5982781246862	-1.05059080356605\\
  };
\addlegendentry{LMMSE naive}

\addplot [PAM16_22km,each nth point=1]
  table[row sep=crcr]{%
  -43.3261079685733	-0.0012320308495954\\
  -38.3261079685733	-0.00139682706945801\\
  -33.3261079685733	-0.00194725989776557\\
  -28.3261079685733	-0.00373930770874501\\
  -23.3261079685733	-0.00949082868786304\\
  -18.3261079685733	-0.0277602241323537\\
  -13.3261079685733	-0.0850011027029718\\
  -8.32610796857331	-0.258728979668941\\
  -3.32610796857331	-0.743271536281399\\
  1.67389203142669	-1.85551301369348\\
  6.6738920314267	-3.70413697506352\\
  11.6738920314267	-5.85485177719679\\
  16.6739159311484	-7.69224368729149\\
  21.6740770375788	-8.85890492536938\\
  26.237652855904	-11.2676756211405\\
  30.9667822044597	-13.0224456743465\\
  35.8083129896253	-14.6685971007734\\
  40.7163180472178	-16.2729982101034\\
  45.6631852884082	-17.8724651177834\\
  50.6324322105946	-19.4902619503379\\
  55.6145896514778	-21.1374243354111\\
  60.6042399730246	-22.8174242612584\\
  65.5982781246862	-24.5287641769762\\
  };

  \addplot [PAM8_22km,each nth point=1]
    table[row sep=crcr]{%
    -43.1654852396304	-0.00136438456985356\\
    -38.1654852396304	-0.00155984242338657\\
    -33.1654852396304	-0.00220336529033989\\
    -28.1654852396304	-0.004282661183158\\
    -23.1654852396304	-0.0109274466688354\\
    -18.1654852396304	-0.0319710124495591\\
    -13.1654852396304	-0.097666362754207\\
    -8.16548523963044	-0.295520214416329\\
    -3.16548523963044	-0.83687023884379\\
    1.83451476036956	-2.03066551951124\\
    6.83451476036956	-3.89670775434917\\
    11.8353201635996	-5.90304453361936\\
    16.8353438159748	-7.4476596267147\\
    21.7292769565594	-9.02745983968889\\
    26.2434682638463	-11.2818362686898\\
    30.9702156883719	-13.0290198317832\\
    35.8102272438587	-14.6744641671178\\
    40.7174152753633	-16.2803902271311\\
    45.6638042939528	-17.8831678366332\\
    50.6327721972275	-19.505384478062\\
    55.6147619131688	-21.1572924000802\\
    60.6043254036177	-22.8415932537428\\
    65.5983124535082	-24.5563148895328\\
    };

\addplot [PAM4_22km,each nth point=1]
  table[row sep=crcr]{%
  -42.7991973564099	0.00158049456157923\\
  -37.7991973564099	0.00133207809453266\\
  -32.7991973564099	0.000503490207734341\\
  -27.7991973564099	-0.0021916161043333\\
  -22.7991973564099	-0.0108310656034109\\
  -17.7991973564099	-0.0381962377544284\\
  -12.7991973564099	-0.123233812447823\\
  -7.7991973564099	-0.375213463881926\\
  -2.7991973564099	-1.03541438642492\\
  2.2008026435901	-2.36990461974224\\
  7.2008026435901	-4.2091770623599\\
  12.2022216878795	-5.90116134660172\\
  17.201772633258	-7.00216855571416\\
  21.7650127155081	-9.33929132498199\\
  26.2756456226393	-11.3262539410905\\
  30.9895098441702	-13.0595022924171\\
  35.8215808749927	-14.7075934493195\\
  40.7241882530319	-16.321809693206\\
  45.6679077888073	-17.9364239667497\\
  50.6352683925081	-19.5725142399774\\
  55.616271463451	-21.2386415577205\\
  60.605219665062	-22.9360248186522\\
  65.5988433086728	-24.6614866303959\\
  };
\addlegendentry{LMMSE}

\addplot [PAM16_btb,each nth point=1]
  table[row sep=crcr]{%
-42.9056684128607	-0.00133359091679602\\
-37.9056684128607	-0.00159394254370043\\
-32.9056684128607	-0.00243094111117723\\
-27.9056684128607	-0.00510110938654173\\
-22.9056684128607	-0.0135764359313684\\
-17.9056684128607	-0.0403346747051655\\
-12.9056684128607	-0.123906297413994\\
-7.90566841286067	-0.377276904845726\\
-2.90566841286068	-1.08500086374356\\
2.09433158713932	-2.71863497740637\\
7.09433158713933	-5.4563073608952\\
12.0945953924374	-8.62919357508384\\
17.094551984994	-11.2996413225131\\
21.8278970529988	-13.8902672770435\\
26.3069934066747	-16.5216477760001\\
31.0026486987015	-19.0237920846824\\
35.8251860047366	-21.4947104640667\\
40.7231501247374	-23.9375300919762\\
45.6651111362329	-26.337860715433\\
50.6323116369013	-28.6699748480682\\
55.6138632114878	-30.8961060611457\\
60.6035245183553	-32.9700055356823\\
65.5977440379636	-34.8513855242972\\
};
\addlegendentry{16 PAM ESR Empirical (WF) L=0km}

\addplot [PAM8_btb,each nth point=1]
  table[row sep=crcr]{%
-42.6912853393617	-0.00148614123820717\\
-37.6912853393617	-0.00179240021410399\\
-32.6912853393617	-0.00276762655217891\\
-27.6912853393617	-0.00586219234370918\\
-22.6912853393617	-0.0156533784121166\\
-17.6912853393617	-0.0464911180754205\\
-12.6912853393617	-0.142482576272441\\
-7.69128533936173	-0.431272892057624\\
-2.69128533936173	-1.22230557371435\\
2.30871466063827	-2.97419078650187\\
7.30871466063827	-5.72695395830605\\
12.3089137298089	-8.68127233003032\\
17.3090229009613	-11.0316404046789\\
21.842000271276	-13.9135833181369\\
26.3170860125427	-16.5182934004798\\
31.0086726516025	-19.0180156383319\\
35.8286482452229	-21.4886221957497\\
40.7251082847058	-23.9318557205737\\
45.6661864051936	-26.3335658631516\\
50.6328918189286	-28.6682716222189\\
55.6141725312246	-30.8988900366533\\
60.6036856816052	-32.9802803473321\\
65.5978224060813	-34.8729299015853\\
};
\addlegendentry{8 PAM ESR Empirical (WF) L=0km}

\addplot [PAM4_btb,each nth point=1]
  table[row sep=crcr]{%
-42.2171942434636	0.00146187285521097\\
-37.2171942434636	0.00107204401127698\\
-32.2171942434636	-0.000184740800682139\\
-27.2171942434636	-0.00419935728182206\\
-22.2171942434636	-0.0169422949835226\\
-17.2171942434636	-0.0570862314506112\\
-12.2171942434636	-0.181469036240229\\
-7.21719424346356	-0.549562986926729\\
-2.21719424346356	-1.51415982915691\\
2.78280575653644	-3.46612035463265\\
7.78280575653644	-6.15657501558961\\
12.7844644153875	-8.68333376292717\\
17.7846994001685	-10.6324895981619\\
21.925145950758	-13.9166719369529\\
26.3716416394041	-16.4757446710929\\
31.0417453445943	-18.9693399829175\\
35.848078041246	-21.4401776984925\\
40.7363661383163	-23.8866573155463\\
45.6726650848425	-26.2952565280914\\
50.6366133308262	-28.6426352443708\\
55.616299891852	-30.8956801577605\\
60.6049096925612	-33.0130199797438\\
65.5985282329856	-34.9584408753189\\
};
\addlegendentry{4 PAM ESR Empirical (WF) L=0km}
\legend{}

\end{axis}
\end{tikzpicture}%

%% file: includes/6_conclusion.tex
\section{Conclusion}
We derived the WF, the optimal affine estimator in the MSE sense, for purely dispersive short-haul fiber-optic  links with SLD. %
Together with a transmit constellation optimization, the WF compensates CD and SLD-nonlinearity and achieves the maximum rates for transmission over $\SI{20}{\kilo\meter}$.
For future work, imperfections in the transceiver, especially impairments of the MZM can be addressed. In addition, the impact of the Kerr nonlinearity could be considered at higher transmit powers. %

%% file: wiener_filter_short_optical_links.bbl
\begin{thebibliography}{10}
\providecommand{\url}[1]{#1}
\csname url@samestyle\endcsname
\providecommand{\newblock}{\relax}
\providecommand{\bibinfo}[2]{#2}
\providecommand{\BIBentrySTDinterwordspacing}{\spaceskip=0pt\relax}
\providecommand{\BIBentryALTinterwordstretchfactor}{4}
\providecommand{\BIBentryALTinterwordspacing}{\spaceskip=\fontdimen2\font plus
\BIBentryALTinterwordstretchfactor\fontdimen3\font minus
  \fontdimen4\font\relax}
\providecommand{\BIBforeignlanguage}[2]{{%
\expandafter\ifx\csname l@#1\endcsname\relax
\typeout{** WARNING: IEEEtran.bst: No hyphenation pattern has been}%
\typeout{** loaded for the language `#1'. Using the pattern for}%
\typeout{** the default language instead.}%
\else
\language=\csname l@#1\endcsname
\fi
#2}}
\providecommand{\BIBdecl}{\relax}
\BIBdecl

\bibitem{8649641}
M.~{Chagnon}, ``Optical {C}ommunications for {S}hort {R}each,'' \emph{J.
  Lightw. Technol.}, vol.~37, no.~8, pp. 1779--1797, April 2019.

\bibitem{7779079}
V.~{Houtsma}, D.~{van Veen}, and E.~{Harstead}, ``Recent {P}rogress on
  {S}tandardization of {N}ext-{G}eneration 25, 50, and {100G} {EPON},''
  \emph{J. Lightw. Technol.}, vol.~35, no.~6, pp. 1228--1234, 2017.

\bibitem{8259239}
K.~{Zhong}, X.~{Zhou}, J.~{Huo}, C.~{Yu}, C.~{Lu}, and A.~P.~T. {Lau},
  ``Digital {S}ignal {P}rocessing for {S}hort-{R}each {O}ptical
  {C}ommunications: {A} {R}eview of {C}urrent {T}echnologies and {F}uture
  {T}rends,'' \emph{J. Lightw. Technol.}, vol.~36, no.~2, pp. 377--400, 2018.

\bibitem{7341665}
S.~{Randel}, D.~{Pilori}, S.~{Chandrasekhar}, G.~{Raybon}, and P.~{Winzer},
  ``100-{G}b/s discrete-multitone transmission over 80-km {SSMF} using
  single-sideband modulation with novel interference-cancellation scheme,'' in
  \emph{2015 European Conference on Optical Communication (ECOC)}, 2015.

\bibitem{AgrawalFourthEdFiberOptics}
G.~Agrawal, \emph{Fiber-Optic Communication Systems: Fourth Edition}, 2010.

\bibitem{Karanov:18}
B.~{Karanov}, M.~{Chagnon}, F.~{Thouin}, T.~A. {Eriksson}, H.~{Bülow},
  D.~{Lavery}, P.~{Bayvel}, and L.~{Schmalen}, ``End-to-{E}nd {D}eep {L}earning
  of {O}ptical {F}iber {C}ommunications,'' \emph{J. Lightw. Technol.}, vol.~36,
  no.~20, pp. 4843--4855, 2018.

\bibitem{abs-1802-00432}
R.~{Ghods}, A.~S. {Lan}, T.~{Goldstein}, and C.~{Studer}, ``Phaselin: Linear
  {P}hase {R}etrieval,'' in \emph{2018 52nd Annual Conference on Information
  Sciences and Systems (CISS)}, 2018, pp. 1--6.

\bibitem{seimetz2009high}
M.~Seimetz, \emph{High-{O}rder {M}odulation for {O}ptical {F}iber
  {T}ransmission}.\hskip 1em plus 0.5em minus 0.4em\relax Springer, 2009, vol.
  143.

\bibitem{gallager2008principles}
R.~Gallager, \emph{Principles of Digital Communication}.\hskip 1em plus 0.5em
  minus 0.4em\relax Cambridge University Press, 2008.

\bibitem{4132995}
S.~{Hranilovic}, ``Minimum-{B}andwidth {O}ptical {I}ntensity {N}yquist
  {P}ulses,'' \emph{{IEEE} Trans. Commun.}, vol.~55, no.~3, pp. 574--583, March
  2007.

\bibitem{kay1993fundamentals}
S.~M. Kay, \emph{Fundamentals of {S}tatistical {S}ignal {P}rocessing}.\hskip
  1em plus 0.5em minus 0.4em\relax Prentice Hall PTR, 1993.

\bibitem{mi_computation_javier_garcia}
\BIBentryALTinterwordspacing
F.~J. Garcia-Gomez, ``Numerically {C}omputing {A}chievable {R}ates of
  {M}emoryless {C}hannels,'' \emph{TUM University Library}, 2019. [Online].
  Available: \url{https://mediatum.ub.tum.de/node?id=1533663}
\BIBentrySTDinterwordspacing

\bibitem{github_repository}
\BIBentryALTinterwordspacing
D.~Plabst, ``{Wiener Filter for Short-Reach Fiber-Optic Links},'' 2020.
  [Online]. Available: \url{https://gitlab.lrz.de/dplabst/imdd_wiener_filter}
\BIBentrySTDinterwordspacing

\end{thebibliography}
